\documentclass[aps,prd,reprint,showpacs,floatfix,amsmath,amssymb,nofootinbib]{revtex4-2}
\usepackage{graphicx,color,dcolumn,booktabs,bm}
\usepackage{txfonts}
\usepackage{overpic}
\usepackage{amssymb}
\usepackage{amsmath}
\usepackage{indentfirst}
\usepackage{feynmf}   %{feynmp}
\usepackage{slashed}  %for Feynman symbols
\usepackage{cases}
\usepackage{color}
\usepackage{multirow}
\usepackage{epstopdf}
\usepackage{ulem}
\usepackage{subfigure}
\usepackage[separate-uncertainty]{siunitx}
\usepackage{verbatim}
\usepackage{lipsum}
\usepackage{tabularx}

\newcommand{\enh}{\nonumber\\&}

\newcommand{\vecq}[2]{\vec{#1}^{\,#2}}

\newcommand{\vo}{\vec{o}\@ifnextchar{^}{\,}{}}

\graphicspath{{PDF/}}

\def\slash#1{\setbox0=\hbox{$#1$}           % set a box for #1
	\dimen0=\wd0                                 % and get its size
	\setbox1=\hbox{/} \dimen1=\wd1               % get size of /
	\ifdim\dimen0>\dimen1                        % #1 is bigger
	\rlap{\hbox to \dimen0{\hfil/\hfil}}      % so center / in box
	#1                                        % and print #1
	\else                                        % / is bigger
	\rlap{\hbox to \dimen1{\hfil$#1$\hfil}}   % so center #1
	/                                         % and print /
	\fi}                                         %
\def\sl#1{\setbox0=\hbox{#1}
	\dimen0=\wd0
	\rlap{\hbox to \dimen0{\hss/\hss}}%
	#1}

\usepackage[colorlinks, citecolor=blue,anchorcolor=red,menucolor=red, linkcolor=cyan,filecolor=red,runcolor=red,urlcolor=blue,frenchlinks=red]{hyperref}

\def\contact{\textnormal{contact}}

 \allowdisplaybreaks

\begin{document}
%\begin{CJK}{GBK}{}

\title{Investigating the two-pion exchange of the double charm $DD^*$ chiral interactions and $T_{cc}$
	}

\author{Hao Xu$^{1}$} \email{xuh2020@nwnu.edu.cn}
\author{Li-Xiang Ren$^{1}$} \email{2023212235@nwnu.edu.cn}

\affiliation{
$^1${{Institute of Theoretical Physics, College of Physics and Electronic Engineering, Northwest Normal University, Lanzhou 730070, China}}	
%$^2${Lanzhou Center for Theoretical Physics, Lanzhou University, Lanzhou 730000, China}\\
}

\begin{abstract}
Under chiral effective field theory, we study the $S$-wave $DD^*$ interactions up to second chiral order at one-loop level, which contain full
contact, one-pion-exchange (OPE) and two-pion-exchange (TPE) contributions. Here, we adopt a new subtraction scheme of the two-particle-reducible contributions, and attempt different
regularization schemes uniquely for the TPE contributions since they are highly divergent on the momentum transfer. Under these schemes we conclude that, in
the $I=0$ channel, the TPE contribution is repulsive, then the competition between this powerful repulsing TPE and
the other two (contact and OPE) results in a quite weak attraction. This explains why $T_{cc}$ has an extremely small binding energy if treated as the
$I=0$ $DD^*$ bound state. This feature resembles that of hidden charm $D\bar{D}^*$ as we investigated in previous work \cite{Xu:2021vsi}, which also interpreted
the extremely near-threshold phenomenon of $X(3872)$. In addition, we also solve the Bethe-Salpeter equation with the chiral interactions for a consistency check.

\begin{comment}
Chiral Lagrangians are constructed by incorporating the chiral symmetry, heavy quark symmetry
as well as proper charge conjugation properties of the heavy mesons. The interacting potentials of the $S$-wave $D\bar{D}^*$ system are calculated
up to the second chiral order at 1-loop level, where complete two-pion exchange interactions are included. We further investigate the behaviors of the potentials in coordinate space,
 and their bound state properties. Our studies indicate that there exists an interacting strength ordering among considered four channels:   
 $\textnormal{str.}[0^+(1^{++})]>\textnormal{str.}[0^-(1^{+-})] > \textnormal{str.}[1^+(1^{+-})] > \textnormal{str.}[1^-(1^{++})]$, where str. stands for the strength of the $D\bar{D}^*$ interaction.
Moreover, we find that $X(3872)$ can be treated as a good candidate of $0^+(1^{++})$ molecular state. $D\bar{D}^*$ also tends to form $0^-(1^{+-})$ and $1^+(1^{+-})$ molecular states, and we expect
future experiments to search for the predicted multiple structures around the $D\bar{D}^*$ mass region.
\end{comment}

\end{abstract}

%\pacs{}

\maketitle

\section{introduction}\label{sec1}

The elaborate study of the heavy hadron interactions is a crucial aspect in the nonperturbative QCD physics. It helps us to uncover the interacting mechanism of
low energy QCD where the fundamental quarks and gluons have been confined, and can also be directly utilized to understand the heavy exotic states and near-threshold phenomena,
even served as a nonperturbative QCD input to other areas. The advantage in the heavy hadron sector is that the emergent symmetry, i.e.~the heavy quark symmetry,
can be applied to benefit our studies.

There has been many kinds of methods to deal with the heavy hadron interactions and related exotic states. For example, the one-boson-exchange (OBE) model is one of
the appropriate dynamical models. This model constructs effective meson interactions by taking into account various meson exchanges ($\pi$, $\sigma$, $\rho$, $\omega$ meson and etc).
For example, Ref.~\cite{Liu:2008fh} successfully applied the OBE model to the $D\bar{D}^*$ and $X(3872)$ study. See Ref.~\cite{Chen:2016qju} for details about the OBE model. 
Other hadron-level approaches have also been widely used such as the contact-range method, local hidden gauge approach, nonrelativistic effective field theory and other effective lagrangian approaches
\cite{AlFiky:2005jd,Guo:2017jvc,Dong:2021juy,Dong:2021bvy,Liu:2024uxn}. There also exists
quark-level approaches, where the interactions and concerned multiquark systems are evaluated based on the quark model \cite{Swanson:2003tb,Yang:2020atz,Ortega:2020tng,Huang:2023jec}. In addition, lattice studies play an indispensable role in the area \cite{Mai:2022eur,Bicudo:2022cqi}.
We also refer to Refs.~\cite{Esposito:2016noz,Ali:2017jda,Lebed:2016hpi,Brambilla:2019esw,Liu:2019zoy,Lucha:2021mwx,Chen:2022asf,Hosaka:2016pey} for other approaches not mentioned.
Among them, chiral field effective theory (ChEFT) is one of the promising frameworks dealing with the heavy hadron interactions, which will be adopted in this work.

 ChEFT was originally proposed to deal
with the nucleon-nucleon interaction \cite{Weinberg:1990rz,Weinberg:1991um}, it has been fully developed in last decades and gives us a deep understanding of
the nuclear force \cite{Epelbaum:2008ga,Meissner:2014lgi,Machleidt:2016rvv,Meissner:2015wva,Epelbaum:2019kcf,Hammer:2019poc,Meng:2022ozq}. Similar to chiral perturbation
theory \cite{Scherer:2012tk},  ChEFT framework respects the chiral symmetry and arranges the low energy nucleon-nucleon interaction in a systematic way, so that the
interaction is classified as direct contact and multi-Goldstone exchange contributions. Here unique
schemes such as Weinberg scheme have to be applied to keep a 
correct power counting. In Weinberg scheme \cite{Weinberg:1990rz,Weinberg:1991um}, with the standard power-counting
rule one calculates the two-particle-irreducible (2PI) diagrams, i.e. the effective potential, then iterates it into the equation (e.g.~the Schr\"{o}dinger or Lippmann-Schwinger equation) to retrieve rest two-particle-reducible (2PR)
contributions.

In a word, two advantages of ChEFT are prominent: the interaction is dealt with in a systematic way, and it manifests as the multi-Goldstone exchanges. So it is promising to apply ChEFT to 
the heavy hadron interactions.
In the heavy hadron sector, ChEFT adopting Weinberg scheme has indeed been developed in recent
years. In Ref.~\cite{Liu:2012vd}, the authors introduced the ChEFT framework to study the doubly-bottomed system $\bar{B}\bar{B}$ up to next-to-leading order (NLO), especially they
analyzed the two-pion-exchange contributions in detail. Based on this Pioneer work~\cite{Liu:2012vd}, the authors in Ref.~\cite{Xu:2017tsr} developed the framework and applied the ChEFT
to the doubly-charmed $DD^*$ system. A molecular state in the $I=0$ channel was predicted then, which just matched the $T^+_{cc}$ state discovered later. The author in Ref.~\cite{Xu:2021vsi} extended to
the hidden-charm $D\bar{D}^*$ system by considering the proper charge conjugation properties of the heavy hadrons. The authors in Ref.~\cite{Liu:2025fhl} systematically studied the
$D^{(*)}\bar{B}^{(*)}$ systems and their coupled-channel effects. Following this framework, other two-heavy hadron systems were also
studied~\cite{Meng:2022ozq,Wang:2018atz,Meng:2019ilv,Wang:2019nvm,Wang:2019ato,Meng:2019nzy,Wang:2020dhf,Chen:2021htr,Chen:2022iil,Zhang:2022pxc,Huang:2024iua}.

Follow our previous work~\cite{Xu:2017tsr},  we continue to investigate the $DD^*$ interactions.    
 $DD^*$ is a typical two heavy hadron system and similar to $NN$, due to their same isospin configurations and similar interactions with the Goldstone bosons, so we can draw
lessons from the researches in the $NN$ sector. Furthermore, because of the additional symmetry, i.e.~heavy quark symmetry, the interactions in the heavy hadron sector can be simplified
to a good approximation.
Specifically, in this work the heavy hadron formalism will be adopted. Therefore, the study of the  $DD^*$ interactions can surely help us to reveal the
non-perturbative QCD dynamics.
On the other hand, due to the observation of  $T_{cc}$ in recent years \cite{LHCb:2021vvq,LHCb:2021auc}, the related $DD^*$ interactions are now surely under the spotlight. 
 
In 2021, $T^+_{cc}$ was observed by the LHCb collaboration in the distribution of the $D^0D^0\pi^+$ invariant mass \cite{LHCb:2021vvq,LHCb:2021auc}, its mass difference referring to the $D^0D^{*+}$ threshold is
only around 300 keV. Its extremely near-threshold nature as well as the exotic $cc\bar{u}\bar{d}$ content implicate $T^+_{cc}$ is a possible $DD^*$ molecular candidate.
In theory, $T_{cc}$ and the $DD^*$ system have been also studied  in various models, such as the
molecular approaches \cite{Ding:2020dio,Meng:2021jnw,Chen:2021vhg,Dong:2021bvy,Yan:2021wdl,Zhao:2021cvg,Wang:2021yld,Cheng:2022qcm,Lin:2022wmj,Abreu:2022sra,Sun:2024wxz},
quark-level approaches \cite{Cheng:2020wxa,Weng:2021hje,Guo:2021yws,Chen:2021cfl,Liu:2023vrk,Ma:2023int,Wu:2024zbx},
lattice simulations\cite{Padmanath:2022cvl,Lyu:2023xro,Meng:2023bmz,Meng:2024kkp,Collins:2024sfi,Francis:2024fwf,Dawid:2024dgy,Gil-Dominguez:2024zmr,Whyte:2024ihh,Prelovsek:2025vbr,PitangaLachini:2025pxr}, as well as the production mechanisms \cite{Li:2023hpk,Chen:2023xhd,Hua:2023zpa,Paryev:2024ors}.
 
So in this work we will study the $DD^*$ interactions  within ChEFT. Besides our earlier work mentioned above~\cite{Xu:2017tsr}, there has been other works discussing the  $DD^*$ interactions
and $T^+_{cc}$ under ChEFT recently. The authors in Ref.~\cite{Wang:2022jop} revisited the $DD^*$ chiral interactions of Ref.~\cite{Xu:2017tsr} by introducing the local momentum-space
regularization. The authors in Ref.~\cite{Zhai:2023ejo} investigated the two-pion-exchange contributions in a general formalism (without the heavy hadron approach). Despite not $DD^*$, authors
in Ref.~\cite{Chacko:2024cax} studied the two-pion exchange of the $B^{(*)}B^{(*)}$ and $B^{(*)}\bar{B}^{(*)}$ systems, where the coupled channels were taken into account.

Notice that in the investigation of the chiral interaction, to properly study the heavy multiquark
states and exotic phenomena, an unitarization procedure is necessary. The unitarization methods have
been widely applied to different areas, e.g. meson-meson interactions, $NN$ interactions, physics at LHC
energies, and quantum gravity. Especially, it achieves successes in the
studies of hadronic bound states and resonances such as $f_0(500)$ and $\Lambda(1405)$, see Refs.~\cite{Oller:1998zr,Oller:1998hw,Oller:2000fj,Oller:2020guq} for details. In this work, due to the lack of the
unitarization we preliminarily discuss the relation
between our calculated interaction and observed $T_{cc}^+$ state.

In this research, we study the  $DD^*$ chiral interactions under ChEFT within heavy hadron formalism, up to NLO at one-loop level, where Weinberg scheme is adopted. Confronting
previous work~\cite{Xu:2017tsr}, here we will further correct the defect of the previous 2PR subtraction method, 
and elaborately explore the regularization schemes uniquely for the two-pion exchange because of the intrinsic difficulty it faces.
We also solve the Bethe-Salpeter (BS) equation to explore the consistency between different iterative equations.
Actually Main conclusions do not change with our new results, but new insights about the $DD^*$ binding force and the $T_{cc}$ state appear.
   
This paper is organized as follows. In Sec.~\ref{SecLagrangian} we briefly review the chiral interactions of the concerned $DD^*$ system,
express the calculated potentials of the individual contact, one-pion-exchange (OPE) and two-pion-exchange (TPE) contributions, then discuss the 2PR reduction scheme and the regularization of the potentials
especially for the TPE. 
In Sec.~\ref{SecResultsSchro}, we present the numerical results along with an elaborate discussion of the TPE contributions under different schemes.
In Sec.~\ref{SecBS}, we iterate the $DD^*$ chiral interactions into the BS equation for comparing.
The last section is the summary.

%%%%%%%%%%%%%%%%%%%%%%%%%%%%%%%%%%
\section{The chiral interactions of the $DD^*$ system} \label{SecLagrangian}
Following the Refs.~\cite{Liu:2012vd,Xu:2017tsr,Xu:2021vsi}, we continue to investigate the $DD^*$ interactions under ChEFT.
This framework arranges the amplitudes with an expansion parameter $\epsilon=p/\Lambda_\chi$ ($p$ stands for a
 small momentum, pion mass, or $D-D^*$ mass splitting). Specifically, we will construct the Lagrangians and interaction potentials by the order
parameter $\epsilon$ with Weinberg's power counting scheme \cite{Weinberg:1990rz,Weinberg:1991um}.
Note that in this work the flavor $SU(2)$ symmetry is adopted.

\subsection{The chiral Lagrangians}
As in Ref.~\cite{Xu:2017tsr}, we will calculate the amplitudes or interaction potentials up to second order $O(\epsilon^2)$
at one-loop level, so the $D^{(*)}D^{(*)}\pi$ Lagrangian and 4$D^{(*)}$ contact Lagrangian at leading order are needed.

Under the flavor $SU(2)$ symmetry, the $DD^*\pi$ lagrangian at $O(\epsilon^1)$ reads
\cite{Burdman:1992gh,Wise:1992hn,Yan:1992gz}:
\begin{eqnarray}\label{LagrangianHpi1}
\mathcal L^{(1)}_{H\phi}&=&-\langle (i v\cdot \partial H)\bar H
\rangle
+\langle H v\cdot \Gamma \bar H \rangle
+g\langle H \slashed u \gamma_5 \bar H\rangle
\nonumber \\  &&-\frac18 \delta \langle H \sigma^{\mu\nu} \bar H \sigma_{\mu\nu} \rangle,
\end{eqnarray}
with the axial vector field $u_\mu={i\over 2} \{\xi^\dagger, \partial_\mu \xi\}$ and chiral connection 
$\Gamma_\mu = {i\over 2} [\xi^\dagger, \partial_\mu\xi]$, where $\xi =\exp(i \phi/2f)$ with
\begin{equation}
\phi=\sqrt2\left(
\begin{array}{cc}
\frac{\pi^0}{\sqrt2}&\pi^+\\
\pi^-&-\frac{\pi^0}{\sqrt2}\\
\end{array}\right).
\end{equation}
The $(D,D^*)$ doublet $H$ is
\begin{eqnarray} \label{Hfield}
&& H=\frac{1+\slashed v}{2}\left(P^*_\mu\gamma^\mu+iP\gamma_5\right),\quad \nonumber \\
&&\bar H=\gamma^0 H^\dag \gamma^0 = \left(P^{*\dag}_\mu \gamma^\mu+iP^\dag \gamma_5\right) \frac{1+\slashed v}{2},\nonumber\\
&& P=(D^0, D^+), \quad P^*_\mu=(D^{*0}, D^{*+})_\mu.
\end{eqnarray}
%$v=(1,0,0,0)$ stands for the 4-velocity of the $H$ field. 

The contact Lagrangian at $O(\epsilon^0)$ is also needed \cite{AlFiky:2005jd,Valderrama:2012jv,Liu:2012vd}:
\begin{eqnarray}\label{Lagrangian4H0}
\mathcal L^{(0)}_{4H}&=&D_{a} \textnormal{Tr} [H \gamma_\mu \bar H ] \textnormal{Tr} [ H
\gamma^\mu\bar H] \nonumber \\  &&+D_{b} \textnormal{Tr} [H \gamma_\mu\gamma_5 \bar H ] \textnormal{Tr} [ H
\gamma^\mu\gamma_5\bar H] \nonumber\\  && +E_{a} \textnormal{Tr} [H
\gamma_\mu\tau^a \bar H ] \textnormal{Tr} [ H \gamma^\mu\tau_a\bar H] \nonumber\\
&&+E_{b} \textnormal{Tr} [H \gamma_\mu\gamma_5\tau^a \bar H ] \textnormal{Tr} [ H
\gamma^\mu\gamma_5\tau_a\bar H],
\end{eqnarray}
where the low energy constants (LECs) $D_a$, $D_b$, $E_a$, $E_b$ should be determined by other methods.
The Lagrangians at $O(\epsilon^2)$, which are needed to renormalize the $O(\epsilon^2)$ loop amplitudes,
can be found in Refs.~\cite{Liu:2012vd,Xu:2017tsr}.

%As in Refs.~\cite{Liu:2012vd,Xu:2017tsr}, 
The adopted Weinberg's power counting scheme \cite{Weinberg:1990rz,Weinberg:1991um} tells us that, with the chiral power counting
we calculate the sum of the two-particle-irreducible (2PI) diagrams first, then iterate it into the Schr\"{o}dinger, Lippmann-Schwinger or BS equation,
where the two-particle-reducible (2PR) contributions can be restored. Here we will calculate the interaction potentials then solve the
Schr\"{o}dinger equation.

\subsection{The diagrams of the $DD^*$ scattering and the $DD^*$ interaction potentials}
The flavor wave functions $\big|I; K \big\rangle$ of the $DD^*$ system with isospin $I=0,1$ and momentum $K$ 
can be written as
\begin{align}\label{IsospinWaveFunction}
&\big|0,0; K \big\rangle = \frac{1}{\sqrt{2}}\bigg( \big| D^0 D^{*+} \big\rangle -  \big| D^+ D^{*0} \big\rangle \bigg),
\nonumber\\
&\big|1,0; K \big\rangle = \frac{1}{\sqrt{2}}\bigg( \big| D^0 D^{*+} \big\rangle +  \big| D^+ D^{*0} \big\rangle \bigg),
\nonumber\\
&\big|1,1; K \big\rangle =  \big| D^+ D^{*+} \big\rangle ,
\nonumber\\
&\big|1,-1; K \big\rangle =  \big| D^0 D^{*0} \big\rangle .
\end{align}

Considering the above flavor wave functions, we can calculate interaction potentials $\mathcal{V}^{I=0,1}$ up to the chiral order $O(\epsilon^2)$ at
one-loop level (NLO). just at this order, the non-trivial TPE contribution emerges.

\begin{figure}[htpb]
	\begin{center}
		\includegraphics[scale=0.45]{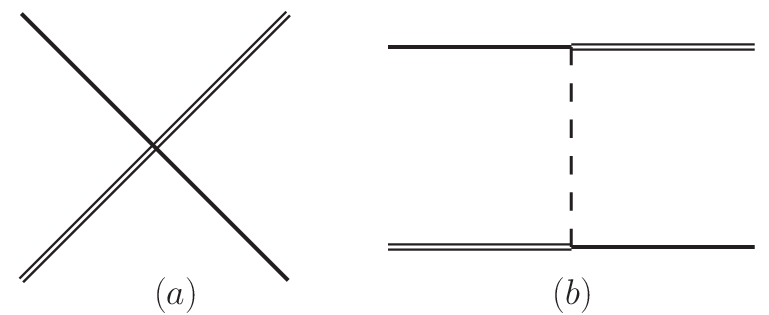}
		\caption{Tree-level diagrams of the $DD^{*}$ system
			at $O(\epsilon^0)$. The solid, double-solid, and dashed lines stand for $D$, $D^{*}$, and $\pi$, respectively.}\label{O0TreeDiagram}
	\end{center}
\end{figure}
We now show the calculated potentials containing all the 2PI contributions.
At $O(\epsilon^0)$ (tree-level), there exists a OPE and a contact 2PI diagram which
are showing in Fig.~\ref{O0TreeDiagram}. With the help of the Lagrangians (\ref{LagrangianHpi1}) and (\ref{Lagrangian4H0})
we obtain their contributions to $\mathcal{V}^{I=0,1}$:

\begin{align}\label{AmplitudeContact0}
\mathcal{V}^{I=0}_{\contact(0)} &= 2 (D_a+D_b-3E_a-3E_b)\varepsilon \cdot \varepsilon^{\,*},
\\
\mathcal{V}^{I=1}_{\contact(0)} &=  2 (D_a-D_b+E_a-E_b)\varepsilon \cdot \varepsilon^{\,*},
\\
\mathcal{V}^{I=0}_{1\pi(0)} &=\frac34 \frac{g^2}{f^2} \frac{1}{q^2-m^2} q\cdot\varepsilon q\cdot\varepsilon^*  ,
\label{AmplitudeOPE00}\\
\mathcal{V}^{I=1}_{1\pi(0)} &=\frac14 \frac{g^2}{f^2} \frac{1}{q^2-m^2} q\cdot\varepsilon q\cdot\varepsilon^*  .
\label{AmplitudeOPE01}
\end{align}
In the above expressions, the subscript (0) of $\mathcal{V}$ stands for the order $O(\epsilon^0)$,
$q$ denotes the momentum transfer between initial and final states, and $m$, $m_1$ and $m_2$ are the masses of $\pi$, $D$ and
$D^*$ respectively. In addition, $\varepsilon$ and $\varepsilon^{\,*}$ are the polarization vectors.

\begin{figure*}[htpb]
	\begin{center}
		\includegraphics[scale=0.99]{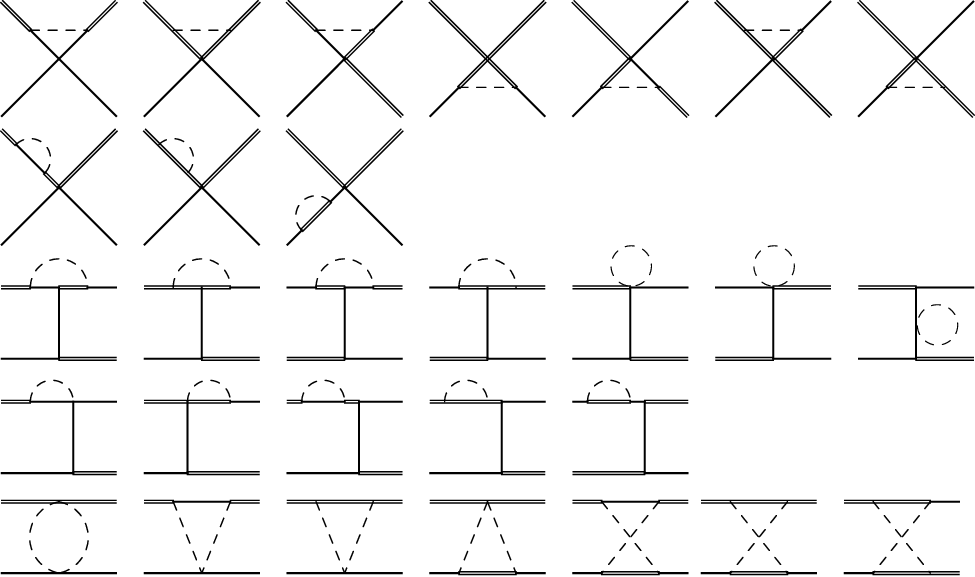}
		\caption{One-loop 2PI diagrams of the $DD^{*}$ system
			at $O(\epsilon^2)$. The solid, double-solid, and dashed lines stand for $D$, $D^{*}$, and $\pi$, respectively.}\label{O21LoopDiagram}
	\end{center}
\end{figure*}

\begin{figure}[htpb]
	\begin{center}
		\includegraphics[scale=0.50]{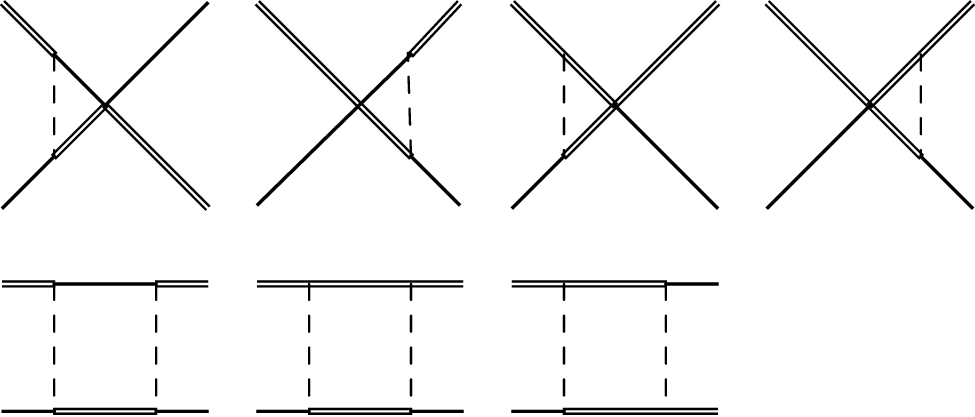}
		\caption{One-loop 2PR diagrams of the $DD^{*}$ system
			at $O(\epsilon^2)$. The solid, double-solid, and dashed lines stand for $D$, $D^{*}$, and $\pi$, respectively.}\label{O21LoopDiagram2PR}
	\end{center}
\end{figure}

For the diagrams at $O(\epsilon^2)$, there appears 1-loop TPE diagrams, as well as 1-loop corrections to
the $O(\epsilon^0)$ contact and OPE diagrams, which are showing in Figs.~\ref{O21LoopDiagram} and \ref{O21LoopDiagram2PR}. As in 
Ref.~\cite{Xu:2017tsr} we ignore the finite parts of the $O(\epsilon^2)$ tree-level amplitudes here due to the lack of their LEC informations,
of course their divergent parts are used to renormalize the $O(\epsilon^2)$ 1-loop diagrams \cite{Xu:2017tsr}. According
to Figs.~\ref{O21LoopDiagram} and \ref{O21LoopDiagram2PR} we now show the calculated $\mathcal{V}$ at $O(\epsilon^2)$ as individual contact, OPE
and TPE contributions:
\begin{widetext}
\begin{align}\label{VContact20}
\mathcal{V}^{I=0}_{\contact(2)} =& \frac{3g^2}{2f^2} \bigg\{  -C_1(d-1)\partial_\omega J^b_{22}(-\delta)-C_1(d-2)\partial_\omega J^b_{22}(0)
-C_1 \partial_\omega J^b_{22}(\delta) -C_1 J^h_{22}(-\delta,\delta) -C_1 J^h_{22}(\delta,-\delta)
\nonumber\\&
 - (dC_2+2C_4)J^g_{22}(-\delta,-\delta) - C_2(d-3)(d-2)J^g_{22}(0,0) - 2C_3J^g_{22}(\delta,\delta) -2C_4J^g_{22}(-\delta,\delta) -2C_4J^g_{22}(\delta,-\delta)
\nonumber\\&
-2C_5(d-3)(d-2)J^h_{22}(0,-\delta) -2C_5(d-3)(d-2)J^h_{22}(-\delta,0)\bigg\}
\varepsilon \cdot \varepsilon^{\,*},
\\\mathcal{V}^{I=1}_{\contact(2)} =&\frac{g^2}{2f^2} \bigg\{ -3C_2(d-1)\partial_\omega J^b_{22}(-\delta) -3C_2(d-2)\partial_\omega J^b_{22}(0)
+3(C_4-C_3)\partial_\omega J^b_{22}(\delta) +(C_4-C_3)J^h_{22}(-\delta,\delta) 
\nonumber\\&
+(C_4-C_3)J^h_{22}(\delta,-\delta) -4C_3J^g_{22}(\delta,\delta) +4C_4J^g_{22}(-\delta,\delta)+4C_4J^g_{22}(\delta,-\delta) -2C_5(d-3)(d-2)J^g_{22}(0,-\delta)
\nonumber\\&
 -2C_5(d-3)(d-2)J^g_{22}(-\delta,0) +\Big[ (2-3d)D_a+(d-2)D_b+(d-6)E_a+5(d-2)E_b\Big]J^g_{22}(-\delta,-\delta)
 \nonumber\\&
  -(d-3)(d-2)(3D_a-D_b-E_a-5E_b)J^g_{22}(0,0)
\bigg\}
\varepsilon \cdot \varepsilon^{\,*},
\\\mathcal{V}^{I=0}_{1\pi(2)} =& \frac{-3}{16f^4} \frac{1}{q^2-m^2}\bigg\{ g^4 \Big[3 (d-1)\partial_\omega J^b_{22}(-\delta) 
+ 3(d-2)\partial_\omega J^b_{22}(0) +3\partial_\omega J^b_{22}(\delta) + (d-3)(d-2)J^g_{22}(0,-\delta) 
 \nonumber\\&
+(d-3)(d-2)J^g_{22}(-\delta,0)
-J^g_{22}(-\delta,\delta) -J^g_{22}(\delta,-\delta)\Big]-8g^2J^c_0 \bigg\} q\cdot\varepsilon q\cdot\varepsilon^*,
\\\mathcal{V}^{I=1}_{1\pi(2)} =& \frac{-1}{16f^4} \frac{1}{q^2-m^2} \bigg\{g^4 \Big[3 (d-1)\partial_\omega J^b_{22}(-\delta) 
+ 3(d-2)\partial_\omega J^b_{22}(0) +3\partial_\omega J^b_{22}(\delta) + (d-3)(d-2)J^g_{22}(0,-\delta) 
\nonumber\\&
+(d-3)(d-2)J^g_{22}(-\delta,0)
-J^g_{22}(-\delta,\delta) -J^g_{22}(\delta,-\delta)\Big]-8g^2J^c_0 \bigg\}q\cdot\varepsilon q\cdot\varepsilon^*,
\\\mathcal{V}^{I=0}_{2\pi(2)} =& \frac{3}{16f^4} \bigg\{ \Big[ -3(d-3)J^B_{22}(-\delta,0)\vec{q}^{\,4} +(d-3)J^R_{22}(-\delta,0)\vecq{q}{4} 
-6(d-3)J^B_{32}(-\delta,0)\vecq{q}{4} +2(d-3)J^R_{32}(-\delta,0)\vecq{q}{4} 
\enh
-3(d-3)J^B_{43}(-\delta,0)\vecq{q}{4} +(d-3)J^R_{43}(-\delta,0)\vecq{q}{4} +6(d-3)J^B_{21}(-\delta,0)\vecq{q}{2} -(d-3)J^R_{21}(-\delta,0)\vecq{q}{2}
+(d-3)J^R_{21}(-\delta,-\delta)\vecq{q}{2} 
\enh
+3(d-3)(2d+1)J^B_{31}(-\delta,0)\vecq{q}{2} +3J^B_{31}(-\delta,\delta)\vecq{q}{2} -(d-3)(2d+1)J^R_{31}(-\delta,0)\vecq{q}{2} -J^R_{31}(-\delta,\delta)\vecq{q}{2}
\enh
+3(d-3)(2d+1)J^B_{42}(-\delta,0)\vecq{q}{2} +3J^B_{42}(-\delta,\delta)\vecq{q}{2} -(d-3)(2d+1)J^R_{42}(-\delta,0)\vecq{q}{2} -J^R_{42}(-\delta,\delta)\vecq{q}{2}
\enh
-3(d-3)(d-2)(d+1)J^B_{41}(-\delta,0) -3(d+1)J^B_{41}(-\delta,\delta) +(d-3)(d-2)(d+1)J^R_{41}(-\delta,0) +(d+1)J^R_{41}(-\delta,\delta)\Big]g^4  
\enh
+ 2\Big[ (d-3)q_0 J^S_{11}(0)\vecq{q}{2} -q_0J^T_{11}(-\delta)\vecq{q}{2} +3(d-3)q_0J^S_{22}(0)\vecq{q}{2} -3q_0J^T_{22}(-\delta)\vecq{q}{2} +2(d-3)J^S_{24}(0)\vecq{q}{2} -2J^T_{24}(-\delta)\vecq{q}{2} 
\enh
+2(d-3)q_0J^S_{32}(0)\vecq{q}{2} -2q_0J^T_{32}(-\delta)\vecq{q}{2} +2(d-3)J^S_{33}(0)\vecq{q}{2} -2J^T_{33}(-\delta)\vecq{q}{2} -(d-3)(d-2)q_0J^S_{21}(0)
-q_0J^S_{21}(\delta) 
\enh
+ (d-1)q_0J^T_{21}(-\delta) -2(d-3)(d-2)q_0J^S_{31}{0} -2q_0J^S_{31}(\delta) + 2(d-1)q_0J^T_{31}(-\delta) -2(d-3)(d-2)J^S_{34}(0) -2J^S_{34}(\delta) 
\enh
+2(d-1)J^T_{34}(-\delta) \Big]g^2 + q_0^2J^F_0 +4q_0^2J^F_{11} +4q_0^2J^F_{21} +4J^F_{22}\bigg\}  
\varepsilon \cdot \varepsilon^{\,*} 
+\frac{2g^2}{16f^4} \bigg\{ \frac32\Big[ -3(d-3)J^B_{22}(\-\delta,0)\vecq{q}{2} + 3J^B_{22}(-\delta,\delta)\vecq{q}{2} 
\enh
+(d-3)J^R_{22}(-\delta,0)\vecq{q}{2} -J^R_{22}(-\delta,\delta)\vecq{q}{2} -6(d-3)J^B_{32}(-\delta,0)\vecq{q}{2} + 6J^B_{32}(-\delta,\delta)\vecq{q}{2} 
+2(d-3)J^R_{32}(-\delta,0)\vecq{q}{2} 
\enh
-2J^R_{32}(-\delta,\delta)\vecq{q}{2} -3(d-3)J^B_{43}(-\delta,0)\vecq{q}{2} +3J^B_{43}(-\delta,\delta)\vecq{q}{2} +(d-3)J^R_{43}(-\delta,0)\vecq{q}{2}
-J^R_{43}(-\delta,\delta)\vecq{q}{2} +6(d-3)J^B_{21}(-\delta,0) 
\enh
-3J^B_{21}(-\delta,\delta) -(d-3)J^R_{21}(-\delta,0) +(d-3)J^R_{21}(-\delta,-\delta) +J^R_{21}(-\delta,\delta) +3(d^2-9)J^B_{31}(-\delta,0) 
-3(d+3)J^B_{31}(-\delta,\delta) 
\enh
-(d^2-9)J^R_{31}(-\delta,0) +(d+3)J^R_{31}(-\delta,\delta) +3(d^2-9)J^B_{42}(-\delta,0) -3(d+3)J^B_{42}(-\delta,\delta) -(d^2-9)J^R_{42}(-\delta,0) 
\enh
+(d+3)J^R_{42}(-\delta,\delta)\Big]g^2 +3(d-3)q_0J^S_{11}(0) -3q_0J^S_{11}(\delta) +9(d-3)q_0J^S_{22}(0) -9q_0J^S_{22}(\delta) + 6(d-3)J^S_{24}(0) -6J^S_{24}(\delta)
\enh
+6(d-3)q_0J^S_{32}(0) -6q_0J^S_{32}(\delta) +6(d-3)J^S_{33}(0) -6J^S_{33}(\delta)
\bigg\} q\cdot\varepsilon q\cdot\varepsilon^*,
\\\mathcal{V}^{I=1}_{2\pi(2)} =&\frac{-1}{16f^4}\bigg\{ \Big[ 
(d-3)J^B_{22}(-\delta,0)\vecq{q}{4} +5(d-3)J^R_{22}(-\delta,0)\vecq{q}{4} +2(d-3)J^B_{32}(-\delta,0)\vecq{q}{4} +10(d-3)J^R_{32}(-\delta,0)\vecq{q}{4} 
\enh
+(d-3)J^B_{43}(-\delta,0)\vecq{q}{4} +5(d-3)J^R_{43}(-\delta,0)\vecq{q}{4} -5(d-3)J^R_{21}(-\delta,0)\vecq{q}{2} -5(d-3)J^R_{21}(-\delta,-\delta)\vecq{q}{2}
\enh
+((5-2d)d+3)J^B_{31}(-\delta,0)\vecq{q}{2} -J^B_{31}(-\delta,\delta)\vecq{q}{2} -5(d-3)(2d+1)J^R_{31}(-\delta,0)\vecq{q}{2} -5J^R_{31}(-\delta,\delta)\vecq{q}{2}
\enh
+((5-2d)d+3)J^B_{42}(-\delta,0)\vecq{q}{2} -J^B_{42}(-\delta,\delta)\vecq{q}{2} -5(d-3)(2d+1)J^R_{42}(-\delta,0)\vecq{q}{2} -5J^R_{42}(-\delta,\delta)\vecq{q}{2}
\enh
+(d-3)(d-2)(d+1)J^B_{41}(-\delta,0) +(d+1)J^B_{41}(-\delta,\delta) +5(d-3)(d-2)(d+1)J^R_{41}(-\delta,0) +5(d+1)J^R_{41}(-\delta,\delta)
\Big]g^4 
\enh
+2\Big[ (d-3)q_0J^S_{11}(0)\vecq{q}{2} -q_0J^T_{11}(-\delta)\vecq{q}{2} +3(d-3)q_0J^S_{22}(0)\vecq{q}{2} -3q_0J^T_{22}(-\delta)\vecq{q}{2} +2(d-3)J^S_{24}(0)\vecq{q}{2}
-2J^T_{24}(-\delta)\vecq{q}{2} 
\enh
+2(d-3)q_0J^S_{32}(0)\vecq{q}{2} -2q_0J^T_{32}(-\delta)\vecq{q}{2} +2(d-3)J^S_{33}(0)\vecq{q}{2} -2J^T_{33}(-\delta)\vecq{q}{2}
-(d-3)(d-2)q_0J^S_{21}(0) -q_0J^S_{21}(\delta) 
\enh
+(d-1)q_0J^T_{21}(-\delta) -2(d-3)(d-2)q_0J^S_{31}(0) -2q_0J^S_{31}(\delta) +2(d-1)q_0J^T_{31}(-\delta) -2(d-3)(d-2)J^S_{34}(0) -2J^S_{34}(\delta) 
\enh
+ 2(d-1)J^T_{34}(-\delta)\Big]g^2 
+ q_0^2J^F_0 +4q_0^2J^F_{11} +4q_0^2J^F_{21} +4J^F_{22}
\bigg\}\varepsilon \cdot \varepsilon^{\,*} 
+\frac{-2g^2}{16f^4}\bigg\{ \frac12\Big[ (d-3)J^B_{22}(-\delta,0)\vecq{q}{2} -J^B_{22}(-\delta,\delta)\vecq{q}{2} 
\enh
+5(d-3)J^R_{22}(-\delta,0)\vecq{q}{2} -5J^R_{22}(-\delta,\delta)\vecq{q}{2} +2(d-3)J^B_{32}(-\delta,0)\vecq{q}{2} -2J^B_{32}(-\delta,\delta)\vecq{q}{2}
+10(d-3)J^R_{32}(-\delta,0)\vecq{q}{2} 
\enh
-10J^R_{32}(-\delta,\delta)\vecq{q}{2} +(d-3)J^B_{43}(-\delta,0)\vecq{q}{2} -J^B_{43}(-\delta,\delta)\vecq{q}{2}
+5(d-3)J^R_{43}(-\delta,0)\vecq{q}{2} -5J^R_{43}(-\delta,\delta)\vecq{q}{2} +J^B_{21}(-\delta,\delta) 
\enh
-5(d-3)J^R_{21}(-\delta,0) -5(d-3)J^R_{21}(-\delta,-\delta) +5J^R_{21}(-\delta,\delta) -(d^2-9)J^B_{31}(-\delta,0) +(d+3)J^B_{31}(-\delta,\delta) 
\enh
-5(d^2-9)J^R_{31}(-\delta,0) +5(d+3)J^R_{31}(-\delta,\delta) -(d^2-9)J^B_{42}(-\delta,0) +(d+3)J^B_{42}(-\delta,\delta) -5(d^2-9)J^R_{42}(-\delta,0)
\enh
+5(d+3)J^R_{42}(-\delta,\delta) \Big]g^2 
+ (d-3)q_0J^S_{11}(0) -q_0J^S_{11}(\delta) +3(d-3)q_0J^S_{22}(0) -3q_0J^S_{22}(\delta) +2(d-3)J^S_{24}(0) -2J^S_{24}(\delta) 
\enh
+2(d-3)q_0J^S_{32}(0) -2q_0J^S_{32}(\delta) +2(d-3)J^S_{33}(0) -2J^S_{33}(\delta)
\bigg\} q\cdot\varepsilon q\cdot\varepsilon^*,\label{V2pi1}
\end{align}
\end{widetext}
where $C_1=D_a+D_b-3(E_a+E_b)$, $C_2=D_a-D_b+E_a-E_b$, $C_3=D_a+E_a$, $C_4=D_b+E_b$, $C_5=D_b-3E_b$. In the above,
$d$ is the space-time dimension originated from the dimensional regularization, $\delta$ is the $D-D^*$ mass splitting, $\mathcal{V}$ relates to the
scattering amplitude $\mathcal{M}$ via $\mathcal{V}=\mathcal{M}/(-\sqrt{\prod 2M_i \prod 2M_f})$. 
In addition, considering the $S$-wave interactions we have $\vec{\varepsilon}\cdot\vecq{\varepsilon}{*} \rightarrowtail 1$
and $\vec{\varepsilon}\cdot \vec{p} \; \vecq{\varepsilon}{*} \cdot \vec{p} \rightarrowtail  \vecq{p}{2}/(d-1)$.

 The definitions of the loop function $J$ are showing
in Eqs.~(\ref{LoopFunction1})-(\ref{LoopFunction2}) (see also Refs.~\cite{Liu:2012vd,Xu:2017tsr}). Also we list the chiral order
of each $J$ in Table \ref{TableJOrder}, so one can conveniently count the orders of Eqs.~(\ref{VContact20})-(\ref{V2pi1}).
They are all collected in Appendix \ref{AppLoopDef}.

Under Fourier transformation we can get the potential $\mathcal{V}(\mathbf{r})$:  
\begin{eqnarray}\label{FourierTransform}
\mathcal{V}(\mathbf{r})=\int \frac{d\mathbf{p}}{(2\pi)^3} \mathcal{V}(\mathbf{p})e^{i\mathbf{p} \cdot\mathbf{r}},
\end{eqnarray}
where $\mathcal{V}(\mathbf{p})$ is a polynomial of $\mathbf{p}$, so the integral will be highly divergent with the increasing of the order.

\subsection{The subtraction of the 2PR contributions and the regularization of the potentials}\label{SecSUbtractRegularization}
To obtain a 2PI effective potential and further substitute it into the equations, we have to subtract the 2PR contributions appearing in all the 2PR 
diagrams, because they violate the power-counting rule and cause pinch singularities in our heavy hadron formalism. 

Now take a closer look at a 2PR diagram, e.g. the box diagram(TPE contribution) in
Fig.~\ref{O21LoopDiagram}. When closing the energy contour integral, we are able to collect two poles in the loop function: the pion pole and the heavy meson pole.
The pion pole belongs to the 2PI part and satisfies the power-counting, whereas the heavy meson pole belongs to the iterated OPE contribution (i.e.~2PR part) and
has an enhanced chiral order which violates the power-counting. Furthermore, the heavy meson poles located in upper and lower planes come
too close, which produces a pinch singularity in the heavy hadron formalism. Therefore we have to preserve only the pion pole.

To collect the pion pole, in our previous work \cite{Xu:2017tsr} we apply the substitution
\begin{align}\label{key}
&\int d^4l \frac{1}{v\cdot l+a+i \varepsilon} \frac{1}{-v\cdot l-a+i \varepsilon} \times \cdots
\enh
\to\int d^4l \frac{1}{v\cdot l+a+i \varepsilon} (-1)\frac{1}{v\cdot l+a+i \varepsilon}\times \cdots
\end{align}
when calculating the loop functions. However, such substitution only subtracts the heavy meson pole in the single channel diagrams (e.g. $DD^*\to D^*D\to DD^*$),
the coupled-channel diagrams (e.g. $DD^*\to D^*D^*\to DD^*$) are intact. So we extend the subtraction to contain all the types of the 2PR diagrams:
\begin{align}\label{key}
&\int d^4l \frac{1}{v\cdot l+a+i \varepsilon} \frac{1}{-v\cdot l-b+i \varepsilon} \times \cdots
\enh
\to\int d^4l \frac{1}{v\cdot l+a+i \varepsilon} (-1)\frac{1}{v\cdot l+b+i \varepsilon}\times \cdots.
\end{align}
Note that although the 2PR parts of the coupled-channel 2PR diagrams avoid the pinch singularities due to a tiny separation of the two heavy meson poles (originated from the mass splitting $\delta$),
their contributions are still abnormally enhanced ($\sim1/\delta$). Therefore the subtractions of these contributions are surely needed. 

After discussing the subtraction of the 2PR contributions, we now turn to the regularization of the calculated 2PI potential. In previous work
\cite{Xu:2017tsr,Xu:2021vsi}, we use a simple Gaussian cutoff exp$(-\vecq{q}{2n}/\Lambda^{2n})$ to regularize the potential. However,
simple use of this regulator leads to two drawbacks in the TPE calculation: an overestimation of the TPE as well as a highly sensitive dependence on the cutoff $\Lambda$.

The reason is that, the calculated TPE potential is a polynomial of momentum $\vec{q}$ (especially contains the terms of high $\vec{q}$ power), this NLO contribution just shows a good convergence within a low $\vec{q}$
region (empirically say, up to $2\pi$ mass, around 300 MeV). When we apply exp$(-\vecq{q}{2n}/\Lambda^{2n})$, the main contribution of
the potential will come from the region $|\vec{q}|\lesssim\Lambda$. So if $\Lambda $ is below 300 MeV, the convergence would be quite good, i.e. the TPE contribution (NLO)
would be relatively small. But in practice, $\Lambda $ is generally chosen to be $500\sim1000$ MeV, it mean that the contributions in the region
$|\vec{q}|\lesssim500$ ($\sim$ $1000$) MeV are all included, which breaks up the convergence and causes TPE abnormally large. Actually, with the increase of $\Lambda$,
the contributions from high $\vec{q}$ reqion will grow rapidly and become dominant, note that these contributions mainly attribute to the terms of high $\vec{q}$ powers
in the potential.

As we discussed above, with a relatively high $\Lambda$, the main contribution of the potential comes from high $\vec{q}$ reqion and the terms of high $\vec{q}$ powers,
it means  the TPE is dominated by short distance contribution. In the coordinate space, It would result in a short distance potential which is singular even than
the contact contribution. Therefore, if insist to use the practically determined $\Lambda$, we need to
suppress the contribution of the TPE in the region $m_{\textnormal{ch}}\lesssim|\vec{q}|\lesssim\Lambda$ (where $m_\textnormal{ch}$ is typically around $2\pi$ mass, about 300 MeV)
in addition to the regulator exp$(-\vecq{q}{2n}/\Lambda^{2n})$. 

So, here we could adopt the following schemes to deal with the above problems in the TPE potential:

$scheme$ I: As we explained above, the chiral expansion is reasonable, while the high $\vec{q}$ contribution is excluded, only when cutoff $\Lambda$ is low,
so we just constrain $\Lambda$ to $0\sim450$ MeV here.
in this case we do not need to introduce additional strategies to suppress the TPE. However, due to the low chosen $\Lambda$, the potential may
 become less attractive with the already fixed LECs, therefore in this scheme we refit the LECs.
 
$scheme$ II:    
In this scheme we do not constrain the $\Lambda$, instead we introduce an regulator $\mathcal{F}_{2\pi}(\vec{q} )$ to the TPE contribution in addition to the
overall regulator exp$(-\vecq{q}{2n}/\Lambda^{2n})$. The purpose of $\mathcal{F}_{2\pi}(\vec{q} )$ is to constrain the highly divergent contribution beyond  
$|\vec{q}|= m_{\textnormal{ch}}$. For $\mathcal{F}_{2\pi}(\vec{q} )$ we use
\begin{align}\label{Regulator2pi}
\mathcal{F}_{2\pi}(\vec{q} )=\left(\frac{\mu_{2\pi}^2}{\vecq{q}{2}+ \mu_{2\pi}^2} \right)^{n_{2\pi}}
\end{align}
with $n_{2\pi}=4$. We adjust $\mu_{2\pi}$ so that the strength of the
TPE outside perturbation region ($|\vec{q}| \gtrsim m_{\textnormal{ch}}$) does not surpass the strength in $|\vec{q}| \lesssim m_{\textnormal{ch}}$.  

$scheme$ III: 
In this scheme we also do not constrain the $\Lambda$, however different from $scheme$ II we adopt another strategy to suppress the TPE. We preserve the 
contribution in the region $|\vec{q}|\lesssim m_{\textnormal{ch}}$, then, we set the momentum potential $\mathcal{V}^{\textnormal{III}}_{2\pi}(\vec{q})$ (superscript III stands for scheme III) to
a constant starting from $|\vec{q}|=m_{\textnormal{ch}}$, i.e., 
\begin{align}\label{Regulator2piIII}
\mathcal{V}^{\textnormal{III}}_{2\pi}(\vec{q})= \left\{
\begin{aligned}
&\mathcal{V}_{2\pi}(\vec{q}) \qquad\qquad\qquad\quad |\vec{q}|\leq m_{\textnormal{ch}},
\\
&\mathcal{V}_{2\pi}(m_{\textnormal{ch}}) \qquad\qquad\qquad |\vec{q}|\geq m_{\textnormal{ch}},
\end{aligned}
\right.
\end{align}
where $\mathcal{V}_{2\pi}(\vec{q})$ is the unmodified TPE contribution. As we discussed above, we choose $m_\textnormal{ch}=300$ MeV. 
This scheme has advantages over the scheme II: it not only suppresses the TPE
in the high $\vec{q}$ region, but lets the TPE having the same ultraviolet behavior as the OPE and contact (they
are both constants in the large $|\vec{q}|$ region), as we will discuss below.

Here we give some comments on the schemes II and III. When we look into
the individual contributions of contact, OPE and TPE in momentum space later in Fig.~\ref{FigVPOrigin}, we can find that with the increase of $\vec{q}$,
the contact and OPE tend to constants
because of the contact terms in them, which means that their sensitivities on $\Lambda$ are same. It further indicates that the two contributions can
adopt a same regularization procedure, i.e. they may be regularized with a same regulator as well as same $\Lambda$. However the TPE is not the case.
The TPE contribution goes to infinity because of the terms of high $\vec{q}$ power, so its ultraviolet behavior is different from that of the contact or
OPE. Consequently the TPE is very sensitive on the cutoff parameter $\Lambda$, so a different regularization or $\Lambda$ may be applied. This is the
reason we introduce the scheme II and scheme III. 

Especially in scheme III, we set the TPE to a constant beyond $m_{\textnormal{ch}}$ within which the expansion is valid, so that all three contributions
have consistent ultraviolet behaviors, based on this we are ready to apply an unified regularization scheme to them. Therefore scheme III
is most promising among the three schemes.
 
At last, note that in all the schemes including I, II and III, we all apply the following overall regulator $\mathcal{F}(\vec{q} )$ to the contact, 
OPE and TPE contributions:
\begin{align}\label{RegulatorI}
&\mathcal{F}(\vec{q} )=\textnormal{exp}(-\vecq{q}{2n}/\Lambda^{2n}).  
\end{align}
Of course we emphasize again that, besides $\mathcal{F}(\vec{q} )$ the TPE are further dealt with using Eq.~(\ref{Regulator2pi}) (scheme II) or
Eq.~(\ref{Regulator2piIII}) (scheme III).
%\begin{align}
%\\&\label{RegulatorII}
%\mathcal{F}(\vec{q} )= \left\{
%\begin{aligned}
%&1 \qquad\qquad\qquad\quad |\vec{q}|\leq m_{\textnormal{ch}},
%\\
%&\textnormal{exp}(-\vecq{q}{2n}/\Lambda^{2n}) \quad\ |\vec{q}|\geq m_{\textnormal{ch}},
%\end{aligned}
%\right. \quad \textnormal{(choice II)}
%\end{align}
%with $n=2$, $m_{\textnormal{ch}}=300$ MeV. Becuse the original regulator (\ref{RegulatorI}) modifies the potential $\mathcal{V}(q)$ when
%$|\vec{q}|=0$, we propose the choice II to maintain the low momentum contribution as much as possible.

After an elaborate numerical analysis, we find that above three schemes give similar potential behaviors and bound state conclusions,
so in the following section we adopt our best choice, i.e.~scheme III, to present the numerical results and discussions. Later we just
briefly mention the results of other two schemes. 

\section{Numerical results of the $DD^*$ potentials} \label{SecResultsSchro}
With the calculated expressions of $\mathcal{V}(q)$, the 2PR subtraction scheme as well as the proposed regularization schemes in Sec.~\ref{SecSUbtractRegularization},
we now present the numerical results and discuss the behaviors of the $DD^*$ potentials with different schemes in detail. In this work we use the input parameters following
Refs~\cite{Xu:2017tsr,Xu:2021vsi}, especially the low energy constants $D^{\textnormal{RSM}}_a=-6.62$, $E^{\textnormal{RSM}}_a=-5.74$, $D^{\textnormal{RSM}}_b=0$ and $E^{\textnormal{RSM}}_b=0$, where RSM means they are obtained using the resonance saturation model. In addition, we use $q^0=\delta$.

\subsection{The original $DD^*$ potentials in momentum space}
\begin{figure}[htpb]
	\begin{center}
		\includegraphics[scale=0.36]{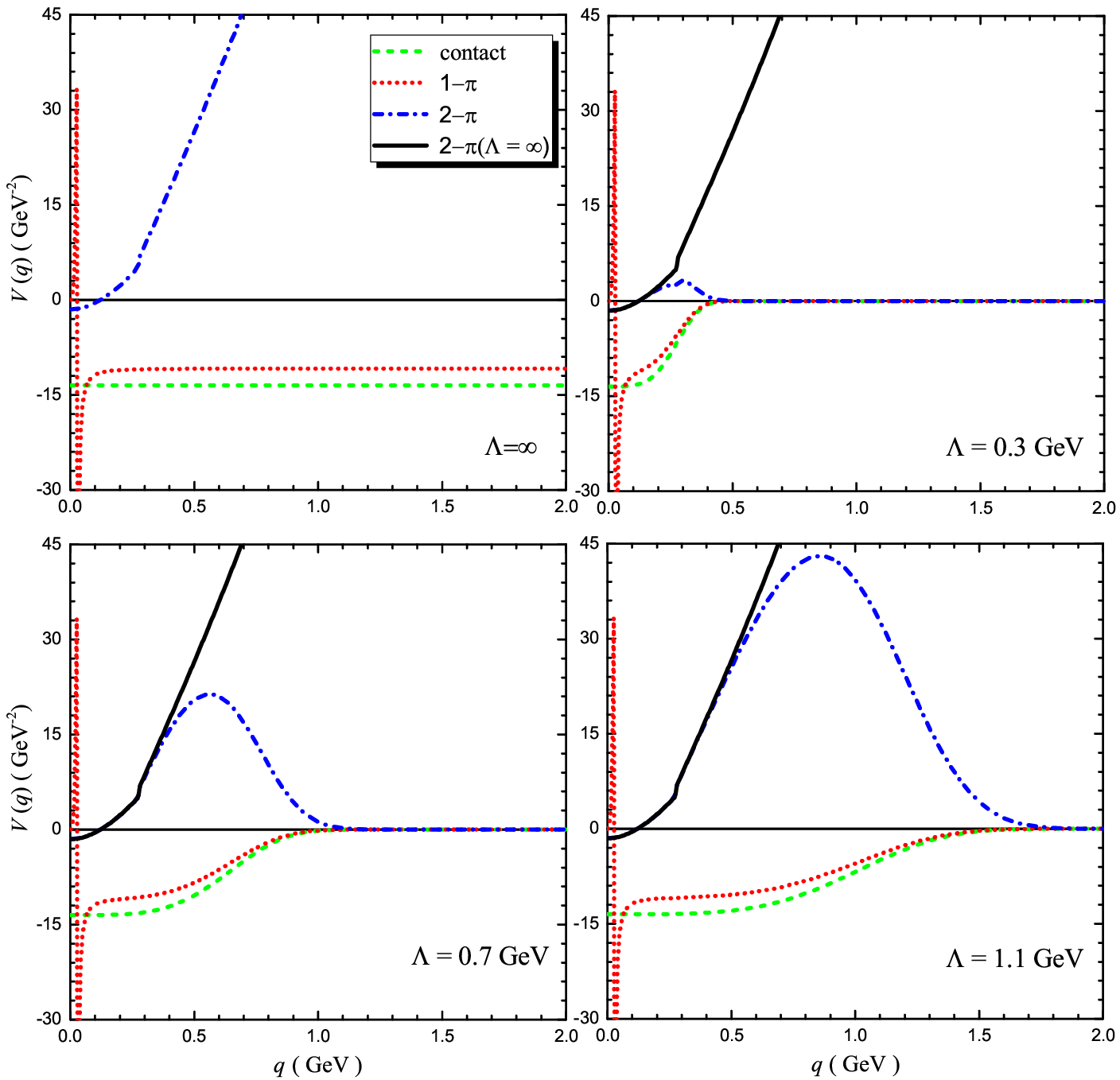}
		\caption{The original $I=0$ $DD^*$ potentials $\mathcal{V}(q)$ in momentum space attached with only overall regulator
	    (\ref{RegulatorI}), varying with cutoff $\Lambda$. Note that $\Lambda=\infty$ just stands for the totally unregulated case. $q=|\vec{q}|$ is the 3-momentum in units of GeV.
		}\label{FigVPOrigin}
	\end{center}
\end{figure}

In this subsection we discuss the potentials in momentum space without any scheme applied, i.e. with only overall
regulator $\mathcal{F}(\vec{q} )$ (\ref{RegulatorI}) attached. We list the results in Fig.~\ref{FigVPOrigin}. Here we consider the
$I=0$ channel. Note that $q=|\vec{q}|$ is the 3-momentum in the Fig~\ref{FigVPOrigin}.

From Fig.~\ref{FigVPOrigin} we can see that, for the results without the regulator ($\Lambda=\infty$), when we restrict ourselves to the low $q$ region (e.g. $q\leq0.3$~GeV) the $O(\epsilon^2)$ TPE is small, which means that the convergence of our chiral expansion series is good. Move
sights to the high $q$ region, the contact and OPE contributions
both tend to finite constants with the increase of $q$, which is reasonable because they have zeroth power of $q$. But the TPE
contribution rapidly goes to divergence with the increase of $q$ because of the terms of high $q$ power in TPE, so its convergence surely breaks up.
As we discussed before, it may indicate that the contact and OPE can be applied with a same regularization scheme while the TPE can not be.  
In conclusion, the high $q$ behavior of the TPE is totally different from that of the contact or OPE.
So we indeed need to adopt a distinctive regularization strategy for the TPE contribution.

We now focus on the results with the overall regulator $\mathcal{F}(\vec{q} )$ turning on. At
$\Lambda=0.3$~GeV, when looking at the general line shape we observe that the TPE contribution is small comparing to the contact and OPE contributions,
which shows a good convergence. As we explained in previous section, a low $\Lambda$ means only the contribution within the low $q$ region is  preserved.
When moving the cutoff to a higher value ($\Lambda=0.7$, $1.1$ GeV) in Fig.~\ref{FigVPOrigin}, we see the suppressed higher region
contributions are gradually recovered. Among them, the TPE is much more sensitive to the $\Lambda$ than the contact and OPE, and it consequently 
becomes abnormally large at a high $\Lambda$, so that its relative magnitude is far beyond the contact and OPE contributions. 
As we dictated in the previous section, such TPE will become a short-distance interaction which is singular
even than the contact. In a word, the TPE should be treated differently in the regularization procedure.

We also mention that in Fig.~\ref{FigVPOrigin}, the regularized TPE
preserves the unregularized TPE potential (black line) in the low $q$ region, while in the high $q$ region they are
suppressed to zero by $\mathcal{F}(\vec{q} )$. So the use of the regulator (\ref{RegulatorI}) is reasonable because the low momentum contributions are well
preserved.

At last, we discuss the general behaviors of all the contributions. From Fig.~\ref{FigVPOrigin} we see the contact and OPE are attractive while the TPE is repulsive.
The judgment of the repulsive TPE is one of our most important conclusions. Note that, it is consistent with the results of the $I=0$ $\bar{B}\bar{B}^*$ system where
$\delta=0$ limit is considered \cite{Wang:2018atz}, and also consistent with the $D^{(*)}\bar{B}^{(*)}$ calculation in Ref.~\cite{Liu:2025fhl}.

\subsection{ $I=0$ $DD^*$ potentials under proposed schemes}
\subsubsection{Scheme III}

Through the above investigations we speculate that a careful treatment of the TPE is needed. So in this subsection we show the $I=0$ results applying 
scheme III, which is the most promising choice among the three schemes as discussed in Sec.~\ref{SecSUbtractRegularization}. 
As we said in this scheme we constrain the TPE using Eq.~(\ref{Regulator2piIII}). This scheme has an advantage over scheme II:
the TPE and other two contributions will have a same high momentum behavior, so that we can confidently use the same regularization scheme (\ref{RegulatorI}) on them.

\begin{figure}[htpb]
	\begin{center}
		\includegraphics[scale=0.355]{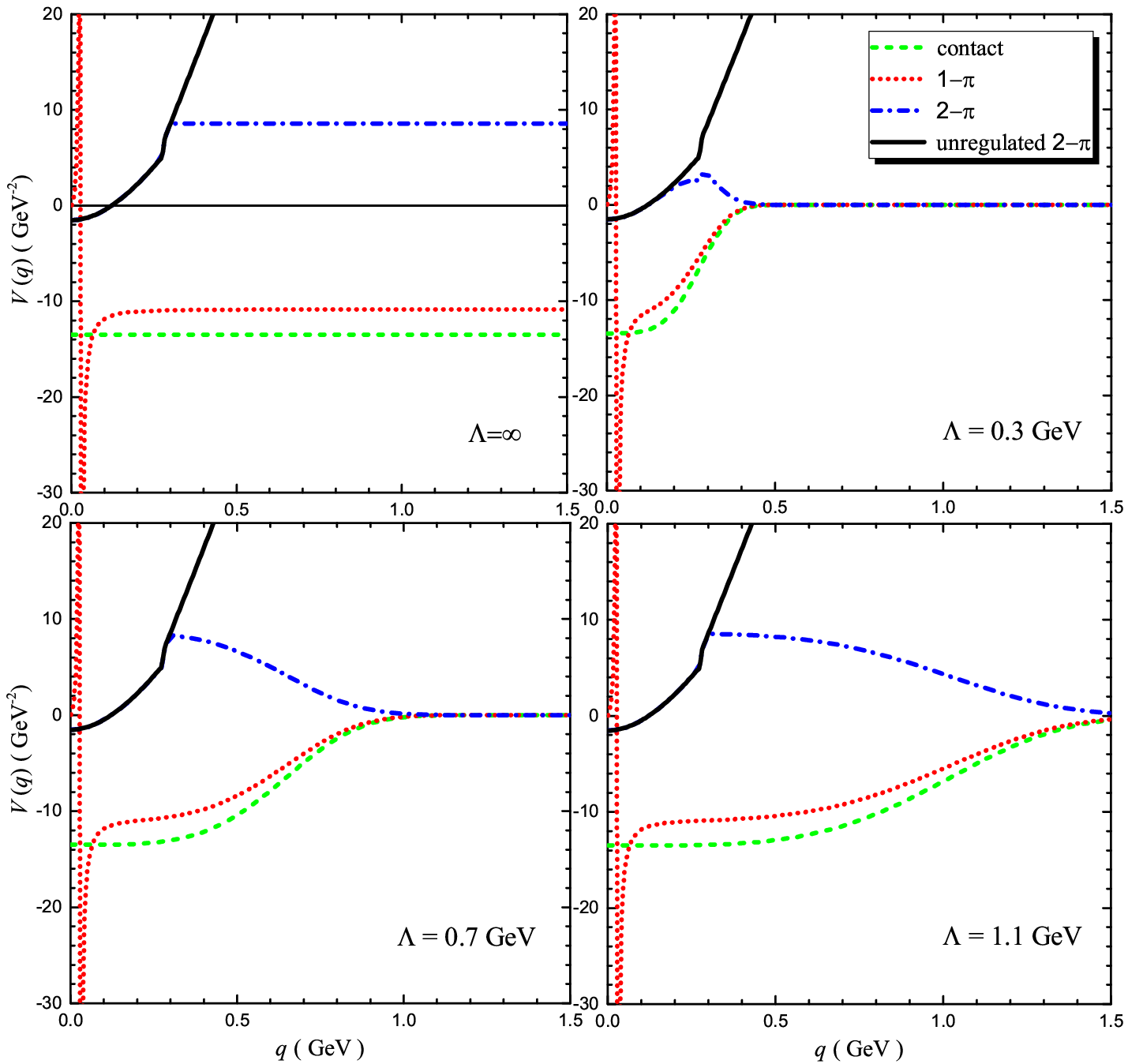}
		\caption{The $I=0$ $DD^*$ potentials $\mathcal{V}(q)$ under scheme III, varying with cutoff $\Lambda$. $q=|\vec{q}|$ is the 3-momentum in units of GeV.
			The ``unregulated $2$-$\pi$" stands for the totally unmodified TPE contribution ($\Lambda=\infty$, and no application of Eq.~(\ref{Regulator2piIII})). 
		}\label{FigVPSchemeIII}
	\end{center}
\end{figure}

We depict the potentials in momentum space varying with $\Lambda$ in Fig.~\ref{FigVPSchemeIII}. From the $\Lambda=\infty$ case we see, in this scheme the TPE
goes to a constant just as other contributions. As we discussed in Sec.~\ref{SecSUbtractRegularization}, in this situation all contributions have a same high $q$ 
(ultraviolet) behavior, so we are able to use the same regularization procedure (Eq.~(\ref{RegulatorI})) on them. It brings a profitable consequence that all three
contributions have highly consistent dependences on the $\Lambda$, as we show in Fig.~\ref{FigVPSchemeIII} with $\Lambda$ varying from 0.3 GeV to 1.1 GeV.
So the TPE is no longer dominant over the other two at high $\Lambda$, and the expansion shows a good convergence.
In other words, it solves the problem discussed in Sec.~\ref{SecSUbtractRegularization}.

\begin{figure}[htpb]
	\begin{center}
		\includegraphics[scale=0.351]{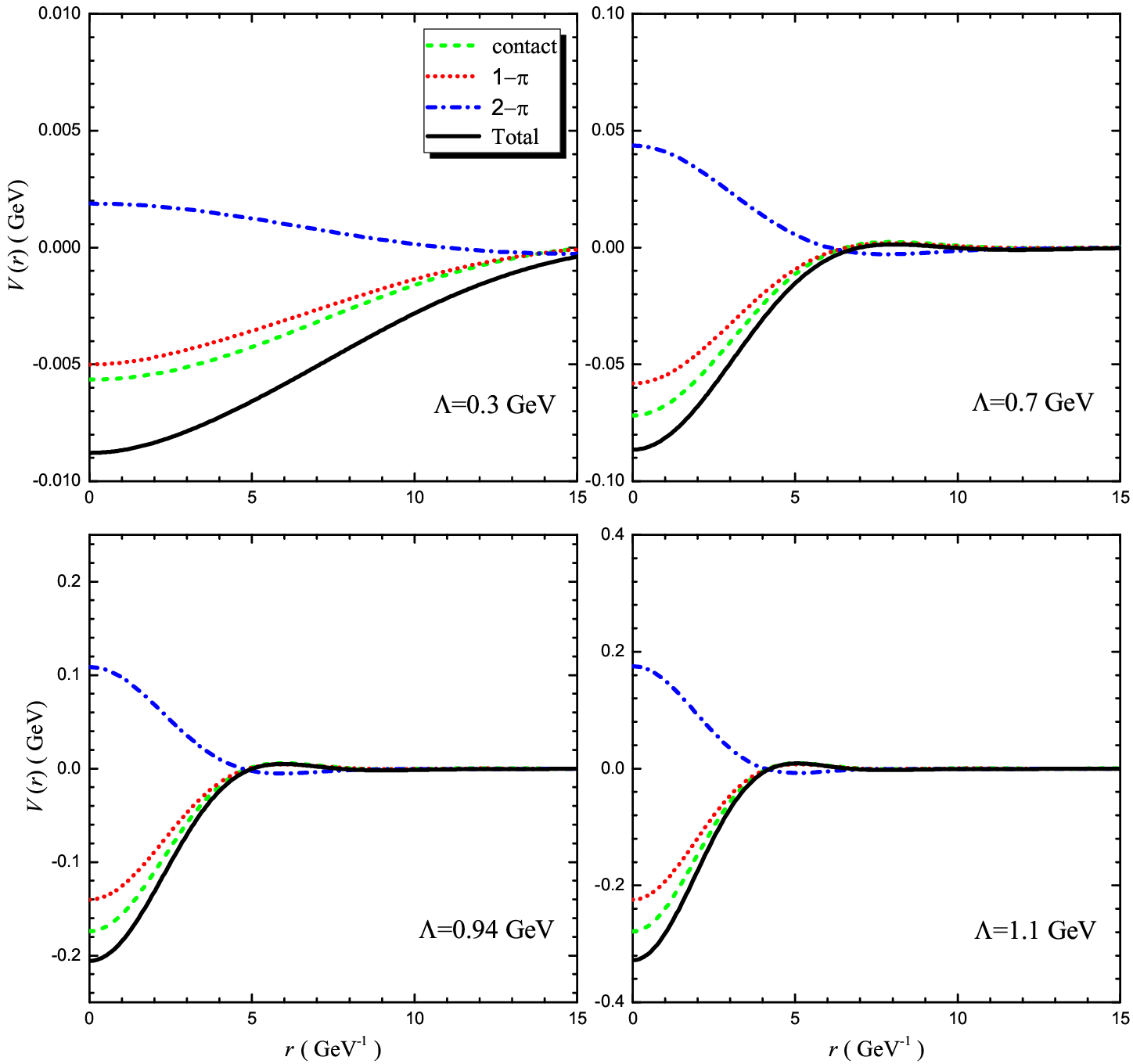}
		\caption{The $I=0$ $DD^*$ potentials $\mathcal{V}(r)$ under scheme III, varying with cutoff $\Lambda$. 
			Note that the potentials at $\Lambda=0.94$ GeV correspond to the observed $T_{cc}$.
		}\label{FigVrSchemeIII}
	\end{center}
\end{figure}

In Fig.~\ref{FigVrSchemeIII} we present the potentials $\mathcal{V}(r)$. 
As we expect the behaviors of the TPE contributions
are consistent with the $\mathcal{V}(q)$ showing in Fig.~\ref{FigVPSchemeIII}: 
the TPE rises synchronously with the contact and OPE on $\Lambda$, and grows smoothly when $\Lambda$ varies from 0.3 GeV to 1.1 GeV, also
it does not get overwhelmingly large at high $\Lambda$. 
In Fig.~\ref{FigVrSchemeIII} at $\Lambda=0.94$ GeV, the potential $\mathcal{V}(r)$ can produce 
the mass of the $T_{cc}$ state when substituted in the Schr\"{o}dinger equation. 

From Fig.~\ref{FigVrSchemeIII} we can see that, the competition between the powerful repulsing TPE and the other two (contact and OPE) leads to a quite weak attraction,
this explains why $T_{cc}$ has a extremely small binding energy if treated as the $I=0$ $DD^*$ bound state. Therefore we believe the repulsive nature of the
TPE in our framework accords with the experimental
observation. On the other hand, if the TPE is attractive, we expect all three attractive contributions should have caused a much deeper $I=0$ $DD^*$ bound state.

The above feature of $T_{cc}$ ($I=0$ $DD^*$) is also very similar to that of $X(3872)$ ($I=0$ $D\bar{D}^*$) presented in our previous work \cite{Xu:2021vsi}.
In Ref.~\cite{Xu:2021vsi}, we find that for $X(3872)$ treated as a $0^+1^{++}$ $D\bar{D}^*$ bound state, its TPE is repulsive while the contact and OPE are attractive, which leads to a small binding energy. In a word, this common feature shared by $X(3872)$ and $T_{cc}$, especially the repulsive TPE, is much responsible for their extremely near-threshold phenomena.

\subsubsection{Brief discussions of scheme I and II}
As we dictated in the end of Sec.~\ref{SecSUbtractRegularization}, we will not present the details of the scheme I and II results here, but give a short discussion. 

In scheme I, other than using an additional regularization on the TPE, we actually restrict cutoff $\Lambda$ to a low value to maintain
the convergence as well as suppress the high $q$ contribution. Here we collect the results in Figs.~\ref{FigVPSchemeI} and \ref{FigVrSchemeI}
in Appendix \ref{AppDiagramSchemeI}. In this scheme we observe
that, because of the adopted low $\Lambda$, the potentials span across large distances, while the magnitudes of the potentials are relatively small.
It means that at these low $\Lambda$, they are very long range potentials.

In scheme II, we apply an additional factor $\mathcal{F}_{2\pi}(\vec{q} )$ (\ref{Regulator2pi})
to constrain the TPE outside the perturbation region ($|\vec{q}| \gtrsim m_{\textnormal{ch}}$). Numerical results show that the TPE contribution in high $q$ region
is largely suppressed comparing to the original highly divergent unmodified TPE just as in scheme III, though under a different strategy.

In a word, in all three schemes the potential behaviors are similar, and their conclusion
about the bound state is
also the same: the near-threshold nature of the $T_{cc}$ state may originate from the competition between the TPE and the other two.

There is a comment about scheme III and II. They all preserve the TPE contribution in the low $q$ region while modify that in the high $q$ region.
We can find scheme II completely cutoffs the high $q$ contribution whereas
scheme III retains it to let the TPE coincide with the contact and OPE at high $q$. Obviously scheme III is better. 
However notice that the TPE at high $q$ are largely cut by the regulator (\ref{RegulatorI}),
so their final potentials present not much discrepancies.

\subsection{$I=1$ $DD^*$ potentials}
Having discuss the $I=0$ $DD^*$ potentials in the different schemes, we now show the $I=1$ potentials. Here we adopt scheme III.

\begin{figure}[htpb]
	\begin{center}
		\includegraphics[scale=0.355]{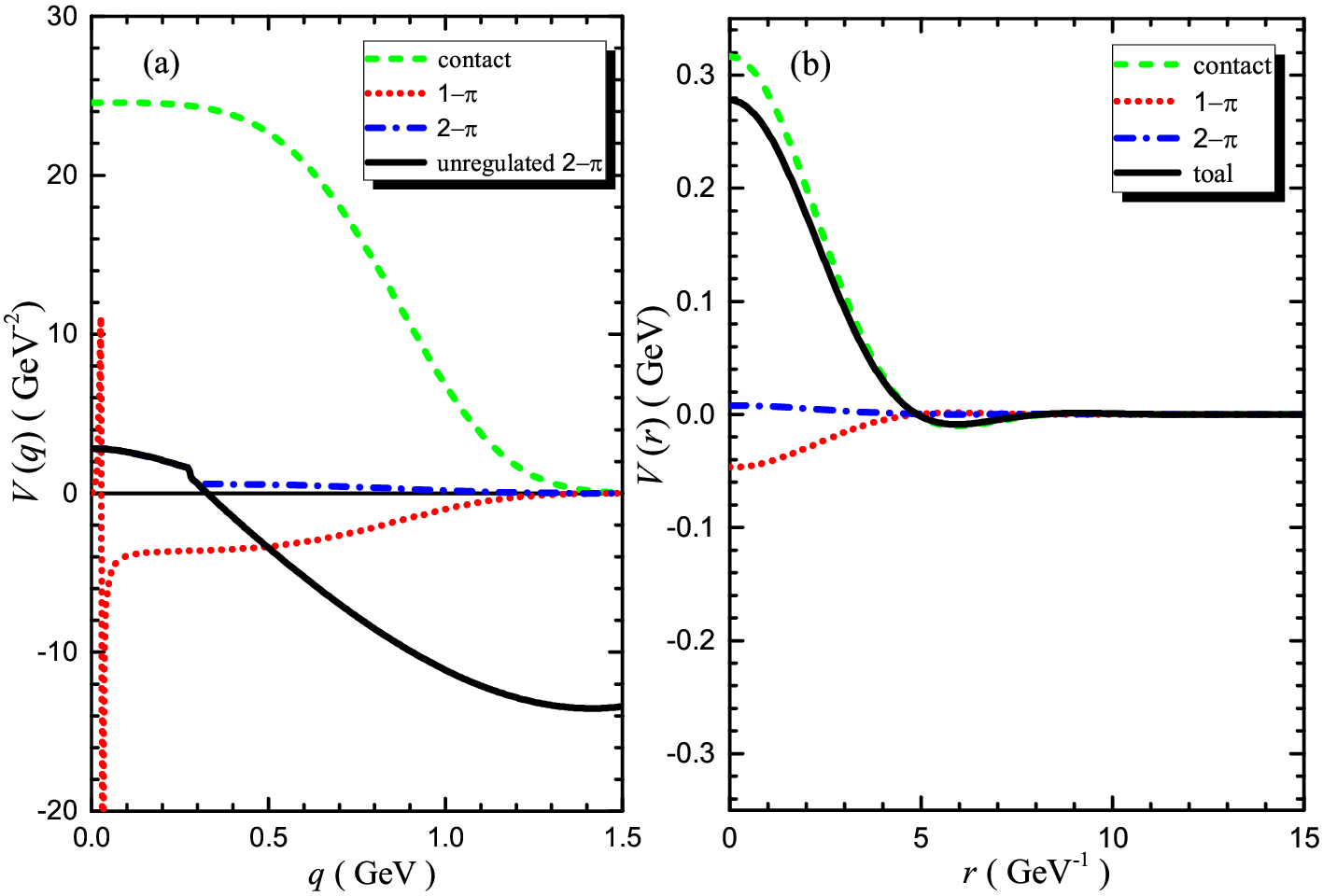}
		\caption{The $I=1$ $DD^*$ potentials under scheme III: (a) the potentials $\mathcal{V}(q)$; (b) the potentials $\mathcal{V}(r)$. $q=|\vec{q}|$ is the 3-momentum in units of GeV.
			The ``unregulated $2$-$\pi$" stands for the totally unmodified TPE contribution ($\Lambda=\infty$, and no application of Eq.~(\ref{Regulator2piIII})). 
		}\label{FigVScheme3I1}
	\end{center}
\end{figure}

We first depict the  $I=1$ $DD^*$ potentials in momentum space in Fig.~\ref{FigVScheme3I1}(a). We can find that the contact contribution is dominant over the small OPE and TPE,
it is also repulsive.

The  $I=1$ $DD^*$ potentials in coordinate space are showing in Fig.~\ref{FigVScheme3I1}(b). Consistent with the $\mathcal{V}(q)$ in Fig.~\ref{FigVScheme3I1}(a), the contact potential 
is dominant and repulsive. So the total potential $\mathcal{V}(r)$ behaves as a strong repulsive force. Indeed, when solving  the Schr\"odinger equation we can not find a bound state
solution within a wide range of $\Lambda$.

\section{Solving the BS equation with the $DD^*$ chiral interactions}\label{SecBS}
In this section we alternatively solve the BS equation to see the consistency between the results of the two equations.
\subsection{The  $DD^*$ BS equation and interaction kernels}
In this subsection we first reduce the BS equation, then give the expressions of the $DD^*$ interaction kernels.

For the $DD^*$ system carrying momentum $K$ and mass $M$, after the center of mass motion is removed the BS wave function is defined as
\begin{align}\label{key}
	\chi^{I, \mu}_{ij} (x,K)
	&=\big\langle 0|T\big[ \mathcal{D}_i(\lambda_2 x)\mathcal{D}^{*\mu}_j(-\lambda_1 x) \big] |I; K \big\rangle
	\enh
	=C^{ij}_{I,I_3} \chi^{I, \mu} (x,K).
\end{align}
Here 
$\mathcal{D}$($\mathcal{D}^{*\mu}$) is the field operator of $D$($D^*$) meson, $x$ is the relative coordinate between $D$ and $D^*$, and we assume $D$($D^*$) possesses
the momentum $k_{1}$($k_{2}$) and mass $m_{1}$($m_{2}$).
The subscript $i(j)=1,2$ is the flavor index, $I$ and $I_3$ are isospin and third component respectively, also $\lambda_{1(2)}=m_{1(2)}/(m_1+m_2)$. 
Note that in above the isospin coefficients $C^{ij}_{I,I_3}$ are determined according to the isospin wave functions (\ref{IsospinWaveFunction}):
\begin{align}\label{BSIsospinCoefficients}
	&C^{12}_{0,0}=\frac{1}{\sqrt{2}}, \quad C^{21}_{0,0}=-\frac{1}{\sqrt{2}},
	\nonumber\\&
	C^{12}_{1,0}=\frac{1}{\sqrt{2}}, \quad C^{21}_{1,0}=\frac{1}{\sqrt{2}},
	\nonumber\\&
	C^{22}_{1,1}=1, \quad C^{11}_{1,-1}=1.
\end{align}

The BS wave function in momentum space $\chi^{I, \mu}_{ij}(k,K)$ is defined via
\begin{align}
	\chi^{I, \mu}_{ij} (x,K)= \int \frac{d^4k}{(2\pi)^4} \chi^{I, \mu}_{ij}(k,K) e^{-ikx},
\end{align}
where $k$ is the relative momentum between $D$ and $D^*$. Using $k$ we also have $k_1=\lambda_1K+k $, $ k_2=\lambda_2K-k$. $\chi^{I, \mu}_{ij}(k,K)$ satisfies 
the following homogeneous integral equation: 
\begin{align}\label{BSEq4DFull}
	\chi^{I, \mu}_{ij} (k,K)=s_D(k_1)s_{D^*}^{\nu\mu}(k_2) \int \frac{d^4l}{(2\pi)^4} \chi^{I, \alpha}_{mn} (l,K) \mathcal{K}_{mnij,\alpha\nu}(l,k,K),
\end{align}
where $s_{D^{(*)}}$ is the propagator for $D^{(*)}$, $\mathcal{K}$ is the interaction kernel. Notice that $l$ is a relative momentum too,
so that $l_1=\lambda_1K+l$ and $l_2=\lambda_2K-l$. Taking into account $\chi^{I, \mu}_{ij}=C^{ij}_{I,I_3} \chi^{I, \mu}$ and $C^{ij}_{I,I_3}$ values~(\ref{BSIsospinCoefficients}),
we  reduce Eq.~(\ref{BSEq4DFull}) to the equation of $\chi^{I, \mu}$:
\begin{align}\label{BSEquation4DI}
	\chi^{\mu} (k)=s_D(k_1)s_{D^*}^{\nu\mu}(k_2) \int \frac{d^4l}{(2\pi)^4} 
	\mathcal{K}_{\alpha\nu}^I(l,k,K)  \chi^{ \alpha} (l) ,
\end{align}
where $\chi^{\mu}$ is the abbreviation of $\chi^{I, \mu}$. The specific $I=0,1$ kernels $\mathcal{K}$ read:
\begin{align}\label{DefKernel}
\mathcal{K}^{I=0}_{\alpha\nu}(l,k,K)&=\mathcal{K}^{D^0D^{*+}\to D^0D^{*+}}(l,k,K)-\mathcal{K}^{D^+D^{*0}\to D^0D^{*+}}(l,k,K),
\nonumber\\
\mathcal{K}^{I=1}_{\alpha\nu}(l,k,K)&=\mathcal{K}^{D^0D^{*+}\to D^0D^{*+}}(l,k,K)+\mathcal{K}^{D^+D^{*0}\to D^0D^{*+}}(l,k,K).
\end{align}
Note that in general a kernel is expressed as a function of the momentum transfer $q$ and $p$: $\mathcal{K}(l,k,K)=\mathcal{K}(p,q,K)$ with relations $q=l_1-k_1$ and $p=l_1-k_2$.

For solving the equations later we decompose the relative momentum $k$ into the longitudinal and transverse parts: $k=k_\ell v+k_t$, $ v\cdot k_t=0$, 
with $v=K/M$ being the four velocity of the $DD^*$ system. Here we consider the instantaneous approximation: $\mathcal{K}(p,q,K)\simeq\mathcal{K}(p_t,q_t,K)$. In this approach we
need to integrate out the longitudinal part of Eq.~(\ref{BSEquation4DI}) to arrive at a three dimension integral equation.
Defining $\chi^\mu(k_t)=\int dk_\ell/2\pi \chi^{ \mu} (k)$, we perform the contour integrals $\int dk_\ell/2\pi$ on both sides of Eq.~(\ref{BSEquation4DI}) and obtain
\begin{align}\label{BSEquation3D}
	\chi^\mu(k_t)=&\int \frac{d^3l_t}{(2\pi)^3} \bigg[
	\frac{-i\big(-g^{\nu\mu}+k_2^\nu k_2^\mu/m_2^2 \big)\Big|_{k_\ell=-\left(\lambda_1M+\omega_1\right)}}{(-2\omega_1)(M+\omega_1+\omega_2)(M+\omega_1-\omega_2)}  
	\nonumber\\&
	+\frac{i\big(-g^{\nu\mu}+k_2^\nu k_2^\mu/m_2^2 \big)\Big|_{k_\ell=\lambda_2M-\omega_2}}{(M-\omega_2+\omega_1)(M-\omega_2-\omega_1)(2\omega_2)}  
	\bigg] \mathcal{K}_{\alpha\nu}^I(p_t,q_t,K)
	\enh
	 \times \chi^\alpha(l_t),
\end{align}
with $\omega_i=\sqrt{m_i^2-k_t^2} $.

We consider the $J^P=1^+$ state only, so the BS wave function $\chi^\mu(k_t)$ can be constructed as
\begin{align}\label{DefBSWF}
	\chi^\mu(k_t)= \varepsilon^\mu(K) \varphi(k_t).
\end{align}

In addition, we normalize the obtained BS wave function with
\begin{align}\label{NormalizationCondition}
	i\int \frac{d^4k}{(2\pi)^4}\frac{d^4l}{(2\pi)^4} \bar{\chi}^\mu(k) \Big[\frac{\partial}{\partial K^0}I_{\mu\nu}(k,l,K) \Big]\chi^\nu(l) =2K^0.
\end{align}

According to the 2PI diagrams in Figs.~\ref{O0TreeDiagram} and \ref{O21LoopDiagram}, we now calculate $\mathcal{K}^{I=0,1}$ (defined in Eq.~(\ref{DefKernel})) up to the
 order $O(\epsilon^2)$. Here we show the  $\mathcal{K}$ at order $O(\epsilon^0)$, 
\begin{align}\label{KContact0}
\mathcal{K}^{I=0,\mu\nu}_{\contact(0)} &= -8i m_1 m_2 (D_a+D_b-3E_a-3E_b)g^{\mu\nu},
\\
\mathcal{K}^{I=1,\mu\nu}_{\contact(0)} &=  -8i m_1 m_2 (D_a-D_b+E_a-E_b)g^{\mu\nu},
\\
\mathcal{K}^{I=0,\mu\nu}_{1\pi(0)} &=-3im_1 m_2 \frac{g^2}{f^2} \frac{1}{p^2-m^2} p^\mu p^\nu ,
\label{KOPE00}\\
\mathcal{K}^{I=1,\mu\nu}_{1\pi(0)} &= -im_1 m_2 \frac{g^2}{f^2} \frac{1}{p^2-m^2} p^\mu p^\nu ,
\label{KOPE01}
\end{align}
while $\mathcal{K}$ at order $O(\epsilon^2)$ are collected in Appendix \ref{AppKernel}.

\subsection{The numerical results}

Here we solve the obtained BS equation above numerically. For the regularization of the TPE we adopt scheme III (\ref{Regulator2piIII}). Note that
due to the distinct formalism comparing to the Schr\"odinger equation, the application of scheme III  here is slightly different. For
the terms with the structure $g^{\mu\nu}$ in the kernels $\mathcal{K}$, we just apply Eq.~(\ref{Regulator2piIII}), while for the terms with the structure $q^\mu q^\nu$
(or $p^\mu p^\nu$), we replace $q^\mu q^\nu$ with $q^\mu q^\nu m_{\textnormal{ch}}^2/\vecq{q}{2}$ beyond $|\vec{q}|=m_{\textnormal{ch}}$ for consistency.

After solving the BS equation, we obtain bound state solutions at some typical $\Lambda$ values, which are listed in Table~\ref{TableDDstarBSSolution}.
We can see that the $I=0$ channel has a bound state solution emerging at around $\Lambda=0.79$ GeV, and the bingding energy increases with the variation of 
$\Lambda$. Also, the solution at $\Lambda=0.79$ corresponds to the observed $T_{cc}$ state. While for the $I=1$ channel there does not exist any solution in all $\Lambda$ range. These conclusions are surely consistent with those obtained 
by solving the Schr\"odinger equation in Sec.~\ref{SecResultsSchro}. 
\setcounter{table}{1}
\begin{table}[!htbp]%The best place to locate the table environment is directly after its first reference in text
	\caption{\label{TableDDstarBSSolution}%
		The bound state solutions of the $DD^*$ BS equations in the $I=0,1$ channels. The cutoff parameter $\Lambda$, calculated mass $M$, and binding energy $E$
		 relative to the $D^0D^{*+}$ threshold are in units of GeV, MeV, and MeV, respectively. Note that the solution at $\Lambda=0.79$ GeV corresponds to the observed $T_{cc}$ state.
	}
 	\begin{ruledtabular}
		\begin{tabular}{ccccc}
& \multicolumn{2}{c}{ $I=0$  } &\multicolumn{2}{c}{  $I=1$ }  \\
\colrule
$\Lambda$	&  $M$   &    $E$    &    $M$   &    $E$   \\
\colrule
$0.5$	&  $-$   &    $-$    &    $-$   &    $-$   \\	
$0.6$	&  $-$   &    $-$  &    $-$   &    $-$    \\
$0.79$	&  $3874.7$   &    $0.36$  &    $-$   &    $-$    \\
$0.8$	&  $3874.2$   &    $0.89$  &      $-$   &    $-$    \\
$0.9$	&  $3869.9$   &    $5.19$  &      $-$   &    $-$  \\
$1.0$	&  $3860.8$   &    $14.29$  &     $-$   &    $-$  \\
$1.1$	&  $3845.8$   &      $29.29$   &    $-$     &    $-$  \\
$1.3$	&  $3796.8$   &    $78.90$   &    $-$   &    $-$   \\
$1.5$	&  $3705.8$   &    $169.29$  &     $-$   &    $-$    \\
		\end{tabular}
	\end{ruledtabular}
\end{table}

\begin{figure}[htpb]
	\begin{center}
		\includegraphics[scale=0.31]{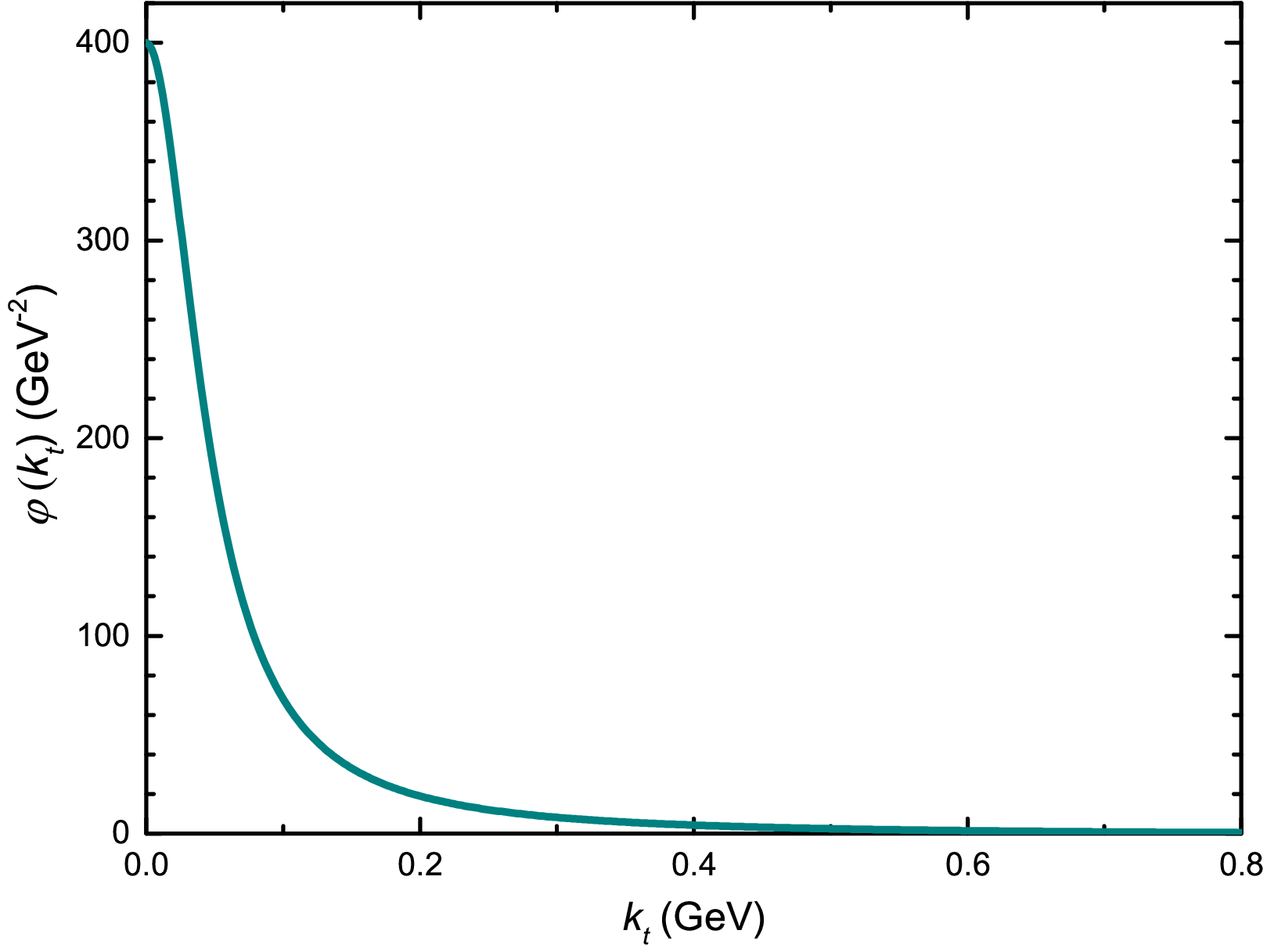}
		\caption{The obtained $I=0$ $DD^*$ BS wave function $\varphi(k_t)$ which corresponds to the solution at $\Lambda=0.79$ GeV, i.e.~the $T_{cc}$ state.
			 $k_t$ denotes  the 3-momentum $|\vec{k}_t|$ in units of GeV. 
		}\label{FigBSWF}
	\end{center}
\end{figure}

Here we also depict the normalized scalar BS wave function $\varphi(k_t)$ of Eq.~(\ref{DefBSWF}) in Fig.~\ref{FigBSWF}. From Fig.~\ref{FigBSWF}, we can find
that the behavior of $\varphi(k_t)$ accords with that of the typical bound state wave function.

\section{Summary} \label{SecS}

In this work, we studied the $DD^*$ interactions under chiral effective field theory (ChEFT). Within heavy hadron formalism, we calculate the $DD^*$ potentials
up to order $O(\epsilon^2)$ (NLO) at 1-loop level. Comparing to our previous work \cite{Xu:2017tsr}, we adopt a new 2PR subtraction scheme for the 2PR digrams,
and discuss multiple regularization strategies uniquely for the TPE. We elaborately investigate the potential behaviors under the proposed regularizations
of the TPE. At last we solve the Schr\"odinger equations with the obtained potentials, and search for the bound state solutions as well as their relations to
the experimental findings. In addition, we also solve the BS equations of the $DD^*$ interactions for a consistency check.

Specifically, among the investigation, our improvements of the framework are as follows. First, due to the intermediate $D^*D^*$ states as the coupled channel appearing in some 
2PR diagrams, we apply a proper 2PR subtraction method to remove the additional poles brought by the coupled channels,
so that all the heavy meson poles are just subtracted. Second, because the TPE contribution is highly divergent on the
momentum transfer $q$ far more than the contact and OPE, we believe a distinctive regularization may be imposed on the TPE. So in this work we introduce
schemes to additionally deal with the TPE regularization. The proposed schemes can properly tackle the TPE in the high $q$ region, so that the TPE and the other two
(contact and OPE) all exhibit consistent and unified behaviors.

Here we list our main conclusions and key findings. We find that, there exists a bound state solution in the $I=0$ channel, while no solution can be found
in the $I=1$ channel, so the observed $T_{cc}$ state indeed can be treated as the $I=0$ $DD^*$ bound state. However, the interaction detail is most interesting:
For the $I=0$ $T_{cc}$ state the TPE contribution is repulsive, then the competition
between the repulsing TPE and other two contributions leads to a quite  weak attraction. This just explains why the binding energy of $T_{cc}$ is extremely
small.

The above feature of $T_{cc}$ just resembles $X(3872)$. As we indicated in Ref.~\cite{Xu:2021vsi}, small binding energy of $X(3872)$ also comes
from the balance between the repulsing TPE and the other two if it is treated as a $0^+1^{++}$ $D\bar{D}^*$ bound state. In conclusion, the repulsive nature of
the TPE, which is shared by both $T_{cc}$ and $X(3872)$, is much responsible for their extremely near-threshold phenomena.

\section*{Acknowledgments}
We thank Chao Wang for giving the valuable support of the BS equation framework, and thank Zhe Liu for checking the calculations. 
This work is supported by the National Natural Science Foundation of China under Grant No. 12465016 and No. 12005168,
the Natural Science Foundation of Gansu Province (No. 22JR5RA171).

\appendix

\section{Definitions of the loop functions} \label{AppLoopDef}
Following Refs.~\cite{Liu:2012vd,Xu:2017tsr}, we define the loop function $J$ as
\begin{widetext}
	\begin{eqnarray}
	&& i\int\frac{d^D l \mu^{4-D} }{ {(2\pi)}^D }\frac{1}{(l^2-m^2+i\varepsilon)}\equiv J^c_0(m)	,\label{LoopFunction1}
	\\
	&& i\int\frac{d^D l \mu^{4-D} }{ {(2\pi)}^D } \frac{ \{1,~
		l^\alpha,~ l^\alpha l^\beta,~ l^\alpha l^\beta l^\gamma\} }
	{[(+/-)v\cdot l+\omega+i\varepsilon](l^2-m^2+i\varepsilon)}
	\nonumber\\&& \equiv \left\{J^{a/b}_0,~ v^\alpha J^{a/b}_{11}, ~
	v^\alpha v^\beta J^{a/b}_{21}+g^{\alpha\beta}J^{a/b}_{22},~ (g\vee
	v)J^{a/b}_{31}+v^\alpha v^\beta v^\gamma J^{a/b}_{32} \right\} (m,\omega),
	\\&&
	i\int\frac{d^D l \mu^{4-D} }{ {(2\pi)}^D } \frac{ \{1,~
		l^\alpha,~ l^\alpha l^\beta,~ l^\alpha l^\beta l^\gamma\} }
	{(v\cdot l+\omega_1+i\varepsilon)[(+/-)v\cdot
		l+\omega_2+i\varepsilon](l^2-m^2+i\varepsilon) }
	\nonumber\\&&\equiv \left\{J^{g/h}_0,~ v^\alpha J^{g/h}_{11},
	~ v^\alpha v^\beta
	J^{g/h}_{21}+g^{\alpha\beta}J^{g/h}_{22},~ (g\vee
	v)J^{g/h}_{31}+v^\alpha v^\beta v^\gamma J^{g/h}_{32} \right\}
	(m,\omega_1,\omega_2),
	\\&&
	i\int\frac{d^D l \mu^{4-D} }{ {(2\pi)}^D } \frac{ \{1,~
		l^\alpha,~ l^\alpha l^\beta,~ l^\alpha l^\beta l^\gamma\} }
	{(l^2-m_1^2+i\varepsilon)[(q+l)^2-m_2^2+i\varepsilon] }
	\nonumber\\&&\equiv \left\{ J^F_0,~ q^\alpha J^F_{11},~
	q^\alpha q^\beta J^F_{21}+g^{\alpha\beta}J^F_{22},~ (g\vee
	q)J^F_{31}+q^\alpha q^\beta q^\gamma J^F_{32} \right\}(m_1,m_2,q),
	\\&&
	i\int\frac{d^D l \mu^{4-D} }{ {(2\pi)}^D } \frac{ \{1,~
		l^\alpha,~ l^\alpha l^\beta,~ l^\alpha l^\beta l^\gamma,~
		l^\alpha l^\beta l^\gamma l^\delta\} } {[(+/-)v\cdot
		l+\omega+i\varepsilon](l^2-m_1^2+i\varepsilon)[(q+l)^2-m_2^2+i\varepsilon]
	} \nonumber\\&&\equiv \left\{ J^{T/S}_0,~ q^\alpha
	J^{T/S}_{11}+v^\alpha J^{T/S}_{12},~ g^{\alpha \beta}
	J^{T/S}_{21}+q^\alpha q^\beta J^{T/S}_{22}+v^\alpha v^\beta
	J^{T/S}_{23}+(q\vee v)J^{T/S}_{24},\right.  (g\vee
	q)J^{T/S}_{31}+q^\alpha q^\beta q^\gamma J^{T/S}_{32} +(q^2\vee
	v)J^{T/S}_{33} \nonumber\\ && +(g\vee v)J^{T/S}_{34}+(q\vee v^2)J^{T/S}_{35}+v^\alpha v^\beta
	v^\gamma J^T_{36},~ (g\vee g)J^{T/S}_{41}+(g\vee
	q^2)J^{T/S}_{42}+q^\alpha q^\beta q^\gamma q^\delta J^{T/S}_{43} +(g\vee
	v^2)J^{T/S}_{44} + v^\alpha v^\beta v^\gamma v^\delta J^{T/S}_{45} \nonumber\\&&
	+(q^3\vee v) J^{T/S}_{46} +(q^2\vee v^2)J^{T/S}_{47} +(q\vee v^3) J^{T/S}_{48}
	\left. +(g\vee q\vee v) J^{T/S}_{49}\right\}(m_1,m_2,\omega,q),
	\end{eqnarray}
	\begin{eqnarray}
	&&
	i\int\frac{d^D l \mu^{4-D} }{ {(2\pi)}^D } \frac{ \{1,~
		l^\alpha,~ l^\alpha l^\beta,~ l^\alpha l^\beta l^\gamma,~
		l^\alpha l^\beta l^\gamma l^\delta\} } {(v\cdot
		l+\omega_1+i\varepsilon)[(+/-)v\cdot
		l+\omega_2+i\varepsilon](l^2-m_1^2+i\varepsilon)[(q+l)^2-m_2^2+i\varepsilon]
	} \nonumber\\&& \equiv \left\{ J^{R/B}_0,~ q^\alpha
	J^{R/B}_{11}+v^\alpha J^{R/B}_{12},~ g^{\alpha \beta}
	J^{R/B}_{21}+q^\alpha q^\beta J^{R/B}_{22}+v^\alpha v^\beta
	J^{R/B}_{23}+(q\vee v)J^{R/B}_{24},\right.  (g\vee
	q)J^{R/B}_{31}+q^\alpha q^\beta q^\gamma J^{R/B}_{32} +(q^2\vee
	v)J^{R/B}_{33} \nonumber\\ &&  +(g\vee v)J^{R/B}_{34} +(q\vee
	v^2)J^{R/B}_{35}+v^\alpha v^\beta v^\gamma J^{R/B}_{36},
	(g\vee g)J^{R/B}_{41}+(g\vee
	q^2)J^{R/B}_{42}+q^\alpha q^\beta q^\gamma q^\delta J^{R/B}_{43}
	+(g\vee v^2)J^{R/B}_{44} + v^\alpha v^\beta v^\gamma v^\delta
	J^{R/B}_{45} \nonumber\\&& +(q^3\vee v) J^{R/B}_{46}+(q^2\vee v^2)J^{R/B}_{47}
	\left.+(q\vee v^3) J^{R/B}_{48}+(g\vee q\vee v)
	J^{R/B}_{49}
	\right\}(m_1,m_2,\omega_1,\omega_2,q),  \label{LoopFunction2}
	\end{eqnarray}
	with
	\begin{eqnarray}
	&&q \vee v \equiv q^\alpha v^\beta+q^\beta v^\alpha, \quad g \vee
	q \equiv
	g^{\alpha\beta}q^\gamma+g^{\alpha\gamma}q^\beta+g^{\gamma\beta}q^\alpha,
	\quad g \vee v \equiv
	g^{\alpha\beta}v^\gamma+g^{\alpha\gamma}v^\beta+g^{\gamma\beta}v^\alpha,
	\quad \nonumber\\&& q^2 \vee v \equiv q^{\beta } q^{\gamma }
	v^{\alpha }+q^{\alpha }
	q^{\gamma } v^{\beta }+q^{\alpha } q^{\beta } v^{\gamma }, \quad
	q \vee v^2 \equiv q^{\gamma } v^{\alpha }
	v^{\beta }+q^{\beta } v^{\alpha } v^{\gamma }+q^{\alpha } v^{\beta } v^{\gamma },\quad
	\nonumber\\&& g \vee g \equiv g^{\alpha \beta } g^{\gamma \delta
	}+g^{\alpha \delta } g^{\beta \gamma }+g^{\alpha \gamma } g^{\beta
		\delta }, \quad g \vee q^2 \equiv q^{\alpha } q^{\beta } g^{\gamma
		\delta }+q^{\alpha } q^{\delta } g^{\beta \gamma } +q^{\alpha}
	q^{\gamma } g^{\beta \delta }+q^{\gamma } q^{\delta } g^{\alpha
		\beta } +q^{\beta } q^{\delta } g^{\alpha \gamma } +q^{\beta }
	q^{\gamma } g^{\alpha \delta }, \quad \nonumber\\&& g \vee v^2
	\equiv v^{\alpha } v^{\beta } g^{\gamma \delta } +v^{\alpha }
	v^{\delta } g^{\beta \gamma }+v^{\alpha } v^{\gamma } g^{\beta
		\delta }
	+v^{\gamma } v^{\delta } g^{\alpha \beta }+v^{\beta } v^{\delta   } g^{\alpha \gamma }
	+v^{\beta } v^{\gamma } g^{\alpha \delta }, \quad
	\nonumber\\&& q^3\vee v \equiv q^{\beta } q^{\gamma } q^{\delta }
	v^{\alpha }+q^{\alpha } q^{\gamma } q^{\delta} v^{\beta }
	+q^{\alpha } q^{\beta } q^{\delta } v^{\gamma }+q^{\alpha }
	q^{\beta } q^{\gamma } v^{\delta } ,\quad q\vee v^3 \equiv
	q^{\delta } v^{\alpha } v^{\beta } v^{\gamma }+q^{\gamma }
	v^{\alpha } v^{\beta } v^{\delta } +q^{\beta } v^{\alpha }
	v^{\gamma } v^{\delta }+q^{\alpha } v^{\beta } v^{\gamma }
	v^{\delta }, \nonumber\\&& q^2 \vee v^2 \equiv q^{\gamma }
	q^{\delta } v^{\alpha } v^{\beta }+q^{\beta } q^{\delta }
	v^{\alpha } v^{\gamma } +q^{\alpha } q^{\delta } v^{\beta }
	v^{\gamma }+q^{\beta } q^{\gamma } v^{\alpha } v^{\delta }
	+q^{\alpha } q^{\gamma } v^{\beta }  v^{\delta }+q^{\alpha }
	q^{\beta } v^{\gamma } v^{\delta }, \nonumber\\&& g\vee q \vee v
	\equiv q^{\beta } v^{\alpha } g^{\gamma \delta }+q^{\alpha }
	v^{\beta } g^{\gamma \delta } +q^{\delta } v^{\alpha } g^{\beta
		\gamma }+q^{\gamma } v^{\alpha } g^{\beta \delta }+q^{\alpha }
	v^{\delta } g^{\beta \gamma } +q^{\alpha } v^{\gamma } g^{\beta
		\delta}+q^{\delta } v^{\gamma } g^{\alpha \beta }+q^{\delta }
	v^{\beta } g^{\alpha \gamma } +q^{\gamma } v^{\delta } g^{\alpha
		\beta  } \nonumber\\&& \qquad\qquad\quad+q^{\gamma } v^{\beta }
	g^{\alpha \delta }+q^{\beta } v^{\delta } g^{\alpha \gamma }
	+q^{\beta } v^{\gamma } g^{\alpha \delta }.
	\end{eqnarray}
\end{widetext}

In the above, $J^b$ is related to $J^a$ via
\begin{eqnarray}
&& J^b_0=J^a_0, \quad J^b_{11}=-J^a_{11}, \quad J^b_{21}=J^a_{21}, \quad J^b_{22}=J^a_{22}, \nonumber \\
&& J^b_{31}=-J^a_{31}, \quad J^b_{32}=-J^a_{32}.
\end{eqnarray}
$J^g$ and $J^h$ can be reduced to
\begin{eqnarray}
&& J^g(\omega_1,\omega_2) =  \frac{1}{\omega_2-\omega_1} \left[ J^a(\omega_1)-J^a(\omega_2)\right],  \\
&& J^h(\omega_1,\omega_2) =  \frac{1}{\omega_2+\omega_1} \left[ J^a(\omega_1)+J^b(\omega_2)\right].
\end{eqnarray}
$J^S$ is related to $J^T$ via
\begin{eqnarray}
&& J^S_0(v \cdot q) = J^T_0(-v \cdot q) , \quad J^S_{11}(v\cdot q)=J^T_{11}(-v\cdot q), \nonumber \\
&& J^S_{12}(v \cdot q)=-J^T_{12}(-v\cdot q), \quad J^S_{21}=J^T_{21}(-v\cdot q), \nonumber \\
&& J^S_{22}(v \cdot q)=J^T_{22}(-v\cdot q), \quad J^S_{23}(v \cdot q)=J^T_{23}(-v \cdot q ). \nonumber \\
&& J^S_{24}(v \cdot q)=-J^T_{24}(-v\cdot q), \quad J^S_{31}(v \cdot q)=J^T_{31}(-v \cdot q ). \nonumber \\
&& J^S_{32}(v \cdot q)=J^T_{32}(-v\cdot q), \quad J^S_{33}(v \cdot q)=-J^T_{33}(-v \cdot q ). \nonumber \\
&& J^S_{34}(v \cdot q)=-J^T_{34}(-v\cdot q), \quad J^S_{35}(v \cdot q)=J^T_{35}(-v \cdot q ). \nonumber \\
&& J^S_{36}(v \cdot q)=-J^T_{34}(-v\cdot q), \quad J^S_{41}(v \cdot q)=J^T_{41}(-v \cdot q ). \nonumber \\
&& J^S_{42}(v \cdot q)=J^T_{42}(-v\cdot q), \quad J^S_{43}(v \cdot q)=J^T_{43}(-v \cdot q ). \nonumber \\
&& J^S_{44}(v \cdot q)=J^T_{44}(-v\cdot q), \quad J^S_{45}(v \cdot q)=J^T_{45}(-v \cdot q ). \nonumber \\
&& J^S_{46}(v \cdot q)=-J^T_{46}(-v\cdot q), \quad J^S_{47}(v \cdot q)=J^T_{47}(-v \cdot q ). \nonumber \\
&& J^S_{48}(v \cdot q)=-J^T_{48}(-v\cdot q), \quad J^S_{49}(v \cdot q)=-J^T_{49}(-v \cdot q ). \nonumber\\
\end{eqnarray}
$J^R$ and $J^B$ can be reduced to
\begin{eqnarray}
&& J^R(\omega_1,\omega_2) =  \frac{1}{\omega_2-\omega_1} \left[ J^T(\omega_1)-J^T(\omega_2)\right],  \\
&& J^B(\omega_1,\omega_2) =  \frac{1}{\omega_2+\omega_1} \left[ J^T(\omega_1)+J^S(\omega_2)\right].
\end{eqnarray}

Notice that in our main text the loop functions $J^{a/b}_{ij}(m,\omega)$, 
$J^{g/h}_{ij}(m,\omega_1,\omega_2)$, $J^{F}_{ij}(m_1,m_2,q)$, $J^{T/S}_{ij}(m_1,m_2,\omega,q)$ and $J^{R/B}_{ij}(m_1,m_2,\omega_1,\omega_2,q)$
in Eqs.~(\ref{LoopFunction1})-(\ref{LoopFunction2}) are abbreviated as $J^{a/b}_{ij}(\omega)$, $J^{g/h}_{ij}(\omega_1,\omega_2)$, $J^{F}_{ij}$,
$J^{T/S}_{ij}(\omega)$ and $J^{R/B}_{ij}(\omega_1,\omega_2)$, respectively.

Here we also list the chiral order of each $J$ in Table~\ref{TableJOrder} based on the definitions~(\ref{LoopFunction1})-(\ref{LoopFunction2}). 
\setcounter{table}{0}
\renewcommand{\arraystretch}{1.3}
\begin{table*}[!htbp]%The best place to locate the table environment is directly after its first reference in text
	\caption{\label{TableJOrder}%
		The chiral order of each loop function $J$. Note that $n$ stands for the order $O(\epsilon^n)$ of a $J$ function.
	}
\begin{ruledtabular}
\begin{tabular}{ccccccccccccccccccc}			
$J$	& $J^{a/b}_0$  & $J^{a/b}_{11}$ & $J^{a/b}_{21}$ & $J^{a/b}_{22}$ & $J^{a/b}_{31}$ & $J^{a/b}_{32}$ & $J^{g/h}_0$ & $J^{g/h}_{11}$ & $J^{g/h}_{21}$ & $J^{g/h}_{22}$ & $J^{g/h}_{31}$ & $J^{g/h}_{32}$ & $J^{F}_0$    & $J^{F}_{11}$  & $J^{F}_{21}$  & $J^{F}_{22}$ & $J^{F}_{31}$ & $J^{F}_{32}$ \\
$n$	&  $1$         &    $2$         &    $3$         &    $3$         &         $4$    &       $4$      &  $0$        &    $1$         &    $2$         &    $2$         &         $3$    &        $3$     &  $0$         &    $0$        &      $0$      &    $2$       &      $2$     &      $0$     \\
			\colrule		
$J$	& $J^{T/S}_0$  & $J^{T/S}_{11}$ & $J^{T/S}_{12}$ & $J^{T/S}_{21}$ & $J^{T/S}_{22}$ & $J^{T/S}_{23}$ &$J^{T/S}_{24}$& $J^{T/S}_{31}$&  $J^{T/S}_{32}$& $J^{T/S}_{33}$ & $J^{T/S}_{34}$ &  $J^{T/S}_{35}$&$J^{T/S}_{36}$&   $J^c_0$     &               &                 & & \\
$n$	&      $-1$    &      $-1$      &      $0$       &      $1$       &      $-1$      &      $1$       &       $0$    &        $1$    &      $-1$      &      $0$       &      $2$       &      $1$       &      $2$     &     $2$       &               &                 & & \\
			\colrule
$J$	&$J^{T/S}_{41}$& $J^{T/S}_{42}$ &$J^{T/S}_{43}$  & $J^{T/S}_{44}$ & $J^{T/S}_{45}$ & $J^{T/S}_{46}$ &$J^{T/S}_{47}$& $J^{T/S}_{48}$&  $J^{T/S}_{49}$&                &                &                &              &               &               &              &               &         \\
$n$	&      $3$     &       $1$      &        $-1$    &      $3$       &      $3$       &      $0$       &      $1$       &      $2$       &      $2$    &                &                &                &              &               &               &              &               &      \\
			\colrule
$J$	& $J^{R/B}_0$  & $J^{R/B}_{11}$ & $J^{R/B}_{12}$ &  $J^{R/B}_{21}$& $J^{R/B}_{22}$ &  $J^{R/B}_{23}$&$J^{R/B}_{24}$&$J^{R/B}_{31}$ & $J^{R/B}_{32}$ & $J^{R/B}_{33}$ & $J^{R/B}_{34}$ & $J^{R/B}_{35}$ &$J^{R/B}_{36}$&               &               &                 & &   \\
$n$	&      $-2$    &      $-2$      &      $-1$      &      $0$       &      $-2$      &      $0$       &       $-1$   &        $0$    &      $-2$      &      $-1$      &      $1$       &      $0$       &      $1$     &               &               &                 & &      \\
			\colrule
$J$	&$J^{R/B}_{41}$& $J^{R/B}_{42}$ &  $J^{R/B}_{43}$&  $J^{R/B}_{44}$&  $J^{R/B}_{45}$& $J^{R/B}_{46}$ &$J^{R/B}_{47}$&$J^{R/B}_{48}$ &  $J^{R/B}_{49}$&                &                &                &              &               &               &              &               &           \\
$n$	&      $2$     &       $0$      &        $-2$    &      $2$       &      $2$       &      $-1$      &      $0$     &      $1$      &      $1$       &                &                &                &              &               &               &              &               &              \\	
\end{tabular}
\end{ruledtabular}
\end{table*}

\section{The results of the potentials in scheme I}\label{AppDiagramSchemeI}

\begin{figure*}[htpb]
	\centering
	\begin{minipage}[t]{0.47\textwidth}
		\centering
		\includegraphics[scale=0.345]{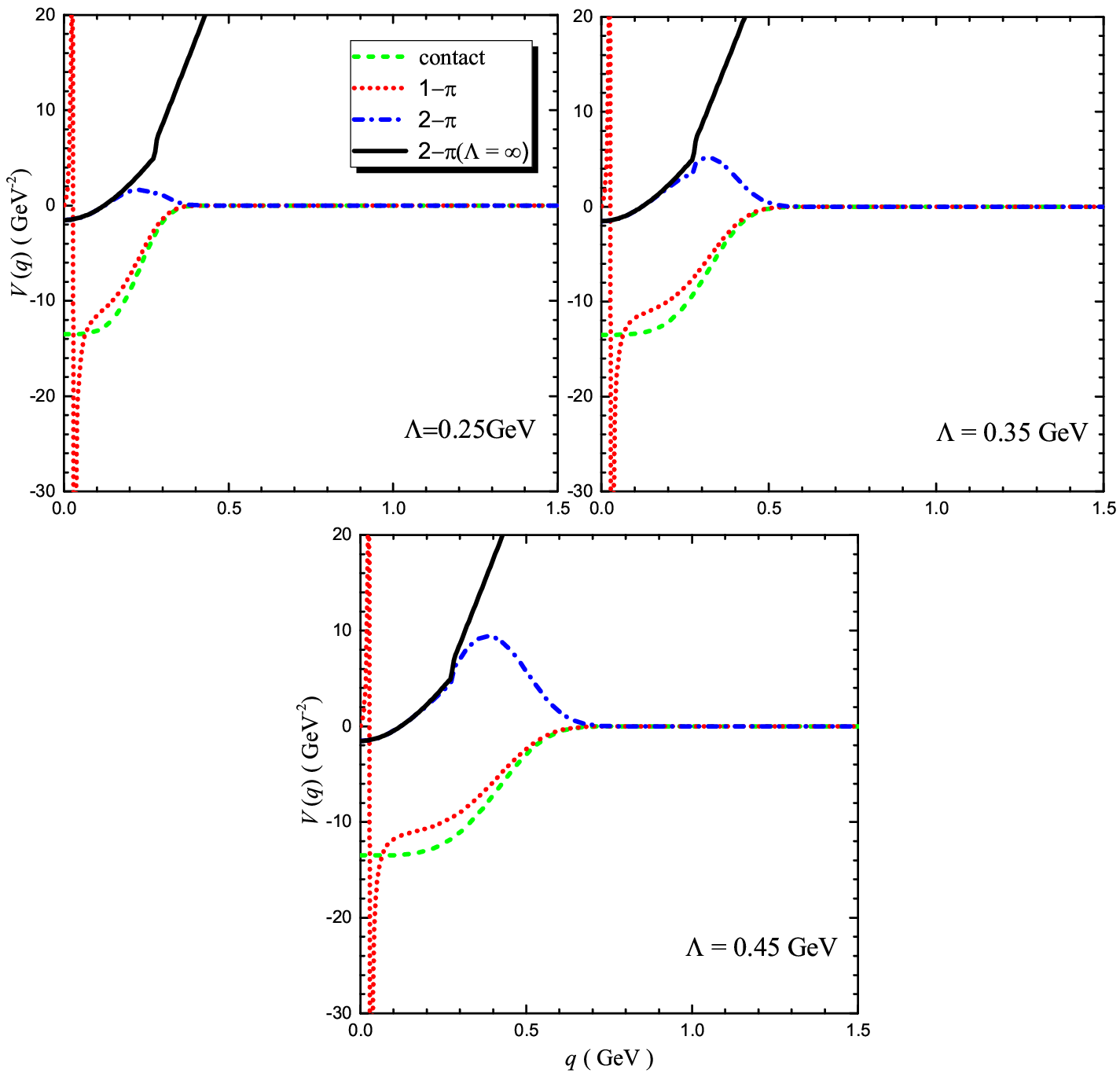}
		\caption{The $I=0$ $DD^*$ potentials $\mathcal{V}(q)$ under scheme I, varying with cutoff $\Lambda$. $q=|\vec{q}|$ is the 3-momentum in units of GeV.
		}\label{FigVPSchemeI}
	\end{minipage}
	\hspace{.14in}
	\begin{minipage}[t]{0.47\textwidth}
		\centering
		\includegraphics[scale=0.345]{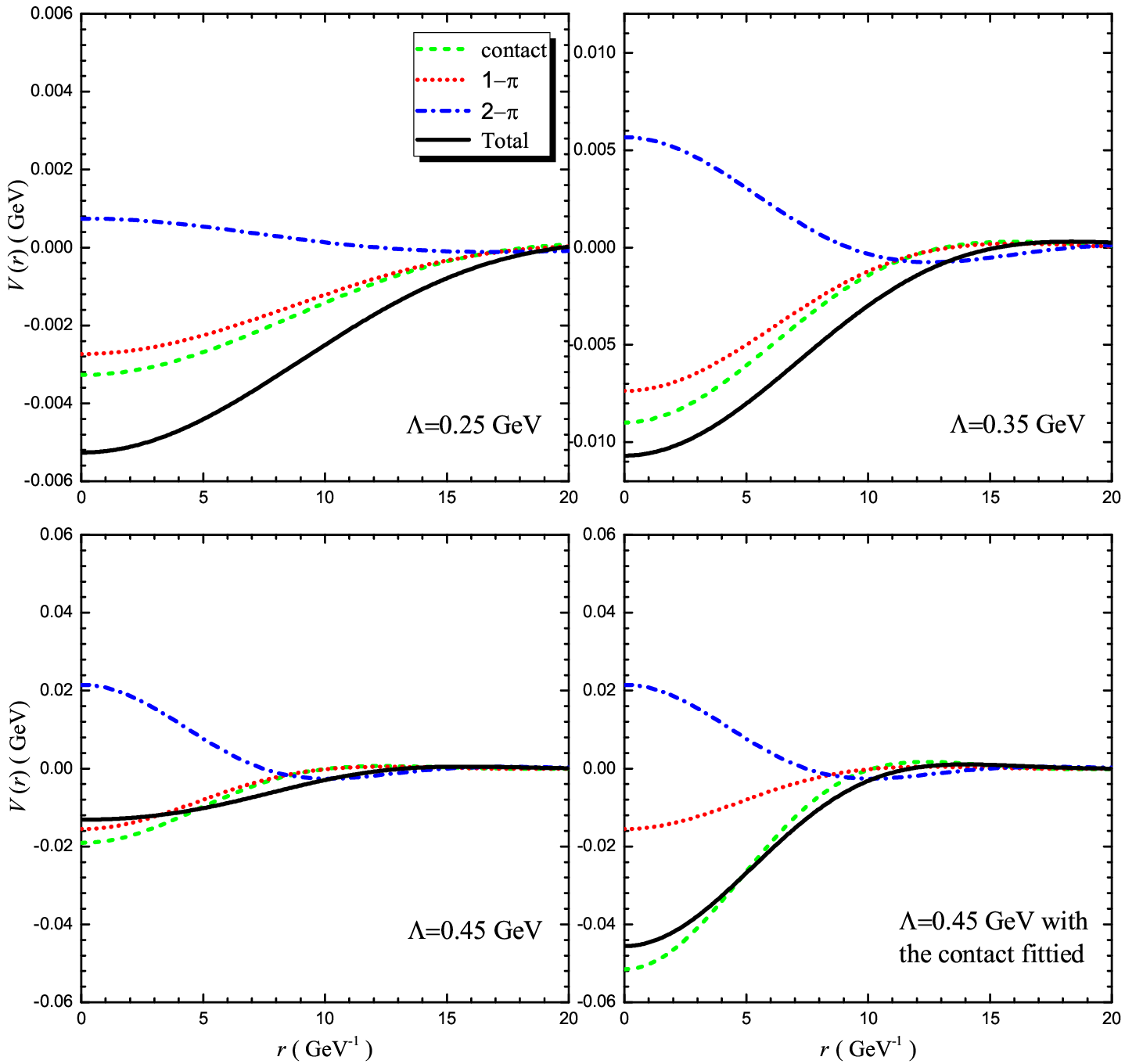}
		\caption{The $I=0$ $DD^*$ potentials $\mathcal{V}(r)$ under scheme I, varying with cutoff $\Lambda$. Note that the LECs of the contact contribution in
			the last diagram are fitted to $T_{cc} $ mass.
		}\label{FigVrSchemeI}
	\end{minipage}
\end{figure*}

Here we show the $I=0$ results under scheme I in Figs.~\ref{FigVPSchemeI} and \ref{FigVrSchemeI}.
We choose three typical $\Lambda$ values: 0.25 GeV, 0.35 GeV and 0.45 GeV.

Due to the insufficient strengths of the potentials at such low $\Lambda$, we can not get the bound state solution. Therefore, 
by solving the Schr\"{o}dinger equation with the $I=0$ $\mathcal{V}(r)$ at $\Lambda=0.45$ GeV, 
we refit the binding energy to the location of the observed $T_{cc}$ state by justifying the LEC $D_a$ and $E_a$. Note that for an illustration,
we do not search for the parameter region of the LECs, but vary them around previous $D^{\textnormal{RSM}}_a$ and $E^{\textnormal{RSM}}_a$ with a same multiplier:
$D_a=cD^{\textnormal{RSM}}_a$, $E_a=cE^{\textnormal{RSM}}_a$. At last we get $c=2.7$ at $\Lambda=0.45$ GeV, and we show the potential $\mathcal{V}(r)$ with the refitted
LECs in the last diagram of Fig.~\ref{FigVrSchemeI}.

\section{The interaction kernels of the $DD^*$ systems at $O(\epsilon^2)$}\label{AppKernel}
Here we list the kernels  $\mathcal{K}^{I=0,1}$ at the second order $O(\epsilon^2)$,

\begin{widetext}
	\begin{align}\label{AmplitudeContact20}
	\mathcal{K}^{I=0,\mu\nu}_{\contact(2)} =& i m_1 m_2\frac{g^2}{f^2} \bigg\{ 6(D_a+D_b-3E_a-3E_b)\Big[ (d-1)\partial_\omega J^b_{22}(-\delta)+ (d-2)\partial_\omega J^b_{22}(0) + \partial_\omega J^b_{22}(\delta) \Big] 
	+ 6(d-2)(d-3)
	\nonumber\\
	&\times (D_a-D_b+E_a-E_b)J^g_{22}(0,0) 
	+ 6\Big[d(D_a-D_b+E_a-E_b)+2(D_b+E_b)\Big]J^g_{22}(-\delta,-\delta) + 12(D_a+E_a)J^g_{22}(\delta,\delta) 
	\nonumber\\
	& + 12(D_b+E_b)J^g_{22}(-\delta,\delta) + 12(D_b+E_b)J^g_{22}(\delta,-\delta) \bigg\} g^{\mu\nu},
	\\
	\mathcal{K}^{I=1,\mu\nu}_{\contact(2)} =& i m_1 m_2\frac{g^2}{f^2} \bigg\{ 6(D_a-D_b+E_a-E_b)\Big[ (d-1)\partial_\omega J^b_{22}(-\delta)+ (d-2)\partial_\omega J^b_{22}(0) + \partial_\omega J^b_{22}(\delta) \Big] 
	+ 2(d-2)(d-3)
	\nonumber\\
	&\times (3D_a-D_b-E_a-5E_b)J^g_{22}(0,0) 
	+ 4(d-2)(d-3)(D_b-3E_b)J^g_{22}(0,-\delta) + 4(d-2)(d-3)(D_b-3E_b)J^g_{22}(-\delta,0)
	\nonumber\\
	&+ 2\Big[(3d-2)D_a-(d-2)D_b+(6-d)E_a+(10-5d)E_b \Big]J^g_{22}(-\delta,-\delta) + 8(D_a+E_a)J^g_{22}(\delta,\delta) 
	\nonumber\\
	& - 8(D_b+E_b)J^g_{22}(-\delta,\delta) -8(D_b+E_b)J^g_{22}(\delta,-\delta) \bigg\} g^{\mu\nu},
	\label{AmplitudeContact21}\\
	\mathcal{K}^{I=0,\mu\nu}_{1\pi(2)} =&im_1 m_2  \frac{1}{p^2-m^2}\bigg\{\frac{g^4}{f^4}
	\bigg[ \frac94\Big((d-1)\partial_\omega J^b_{22}(-\delta)+ (d-2)\partial_\omega J^b_{22}(0) + \partial_\omega J^b_{22}(\delta)\Big) 
	+ \frac34(d-2)(d-3)J^g_{22}(0,-\delta) 
	\nonumber\\
	&+ \frac34(d-2)(d-3)J^g_{22}(-\delta,0) -\frac34 J^g_{22}(-\delta,\delta) -\frac34 J^g_{22}(\delta,-\delta) \bigg] 
	- 6\frac{g^2}{f^4}J^c_0\bigg\} p^\mu p^\nu ,
	\label{AmplitudeOPE20}\\
	\mathcal{K}^{I=1,\mu\nu}_{1\pi(2)} =&im_1 m_2  \frac{1}{p^2-m^2}\bigg\{\frac{g^4}{f^4}
	\bigg[ \frac34\Big((d-1)\partial_\omega J^b_{22}(-\delta)+ (d-2)\partial_\omega J^b_{22}(0) + \partial_\omega J^b_{22}(\delta)\Big) 
	+ \frac14(d-2)(d-3)J^g_{22}(0,-\delta) 
	\nonumber\\
	&+ \frac14(d-2)(d-3)J^g_{22}(-\delta,0) -\frac14 J^g_{22}(-\delta,\delta) -\frac14 J^g_{22}(\delta,-\delta) \bigg] 
	- 2\frac{g^2}{f^4}J^c_0\bigg\} p^\mu p^\nu ,
	\label{AmplitudeOPE21}
	\end{align}
	\begin{align}
\mathcal{K}^{I=0,\mu\nu}_{2\pi(2)} = \mathcal{K}^{I=0,\mu\nu}_{2\pi(2)}(q) +\mathcal{K}^{I=0,\mu\nu}_{2\pi(2)}(p),
	\end{align}
	with
	\begin{align}
	\mathcal{K}^{I=0,\mu\nu}_{2\pi(2)}(q) =&im_1 m_2 \frac{-3}{4f^4} \bigg\{ \Big[
	(d-3)J_{22}^R(-\delta,0) \vecq{q}{4} + 2(d-3) J^R_{32}(-\delta,0)\vecq{q}{4} +(d-3)J^R_{43}(-\delta,0) \vecq{q}{4} -(d-3)J^R_{21}(-\delta,0)\vecq{q}{2}
	\enh 
	-(2d+1)(d-3)J^R_{31}(-\delta,0)\vecq{q}{2} -J^R_{31}(-\delta,\delta)\vecq{q}{2} -(2d+1)(d-3)J^R_{42}(-\delta,0)\vecq{q}{2} -J^R_{42}(-\delta,\delta)\vecq{q}{2}
	\enh
	+(d-3)(d-2)(d+1)J^R_{41}(-\delta,0) + (d+1)J^R_{41}(-\delta,\delta)
	\Big]g^4 
	+ 2\Big[
	(d-3)q_0J^S_{11}(0)\vecq{q}{2} -q_0J^T_{11}(-\delta)\vecq{q}{2} 
	\enh
	 +3(d-3)q_0J^S_{22}(0)\vecq{q}{2} -3q_0J^T_{22}(-\delta)\vecq{q}{2} +2(d-3)J^S_{24}(0)\vecq{q}{2} -2J^T_{24}(-\delta)\vecq{q}{2} + 2(d-3)q_0J^S_{32}(0)\vecq{q}{2}
	 -2q_0J^T_{32}(-\delta)\vecq{q}{2} 
	 \enh
	 +2(d-3)J^S_{33}(0)\vecq{q}{2} -2J^T_{33}(-\delta)\vecq{q}{2} -(d-3)(d-2)q_0J^S_{21}(0) -q_0J^S_{21}(\delta) +(d-1)q_0J^T_{21}(-\delta) 
	 \enh
	 -2(d-3)(d-2)q_0J^S_{31}(0) -2q_0J^S_{31}(\delta) +2(d-1)q_0J^T_{31}(-\delta) -2(d-3)(d-2)J^S_{34}(0) -2J^S_{34}(\delta) +2(d-1)J^T_{34}(-\delta) 
	\Big]g^2 
	\enh
	+ q_0^2J^F_0 +4q_0^2J^F_{11} +4q_0^2J^F_{21} +4J^F_{22}
	\bigg\}g^{\mu\nu}
	+im_1 m_2 \frac{-3g^2}{2f^4} \bigg\{ 
	\frac{-1}{2}\Big[ 
	-(d-3)J^R_{22}(-\delta,0)\vecq{q}{2} +J^R_{22}(-\delta,\delta)\vecq{q}{2} 
	\enh
	-2(d-3)J^R_{32}(-\delta,0)\vecq{q}{2} +2J^R_{32}(-\delta,\delta)\vecq{q}{2} -(d-3)J^R_{43}(-\delta,0)\vecq{q}{2} +J^R_{43}(-\delta,\delta)\vecq{q}{2}
	+(d-3)J^R_{21}(-\delta,0) -J^R_{21}(-\delta,\delta) 
	\enh
	+(d^2-9)J^R_{31}(-\delta,0) -(d+3)J^R_{31}(-\delta,\delta) +(d^2-9)J^R_{42}(-\delta,0) -(d+3)J^R_{42}(-\delta,\delta)
	\Big]g^2 
	+ (d-3)q_0J^S_{11}(0) -q_0J^S_{11}(\delta) 
	\enh
	+3(d-3)q_0J^S_{22}(0) -3q_0J^S_{22}(\delta) +2(d-3)J^S_{24}(0) -2J^S_{24}(\delta) +2(d-3)q_0J^S_{32}(0) -2q_0J^S_{32}(\delta) +2(d-3)J^S_{33}(0) 
	\enh
	-2J^S_{33}(\delta)
	\bigg\}q^\mu q^\nu,
	\\
	\mathcal{K}^{I=0,\mu\nu}_{2\pi(2)}(p) =&im_1 m_2 \frac{-3(d-3)g^4}{4f^4} J^R_{21}(-\delta,-\delta) \Big[\vecq{p}{2}g^{\mu\nu} +p^\mu p^\nu\Big],
	\end{align}
	\begin{align}
\mathcal{K}^{I=1,\mu\nu}_{2\pi(2)} = \mathcal{K}^{I=1,\mu\nu}_{2\pi(2)}(q) +\mathcal{K}^{I=1,\mu\nu}_{2\pi(2)}(p),
\end{align}
with	
	\begin{align}
\mathcal{K}^{I=1,\mu\nu}_{2\pi(2)}(q) =&im_1 m_2 \frac{1}{4f^4} \bigg\{ (-5)\Big[ 
-(d-3)J^R_{22}(-\delta,0)\vecq{q}{4} -2(d-3)J^R_{32}(-\delta,0)\vecq{q}{4} -(d-3)J^R_{43}(-\delta,0)\vecq{q}{4} +(d-3)J^R_{21}(-\delta,0)\vecq{q}{2} 
\enh
+(d-3)(2d+1)J^R_{31}(-\delta,0)\vecq{q}{2} +J^R_{31}(-\delta,\delta)\vecq{q}{2} +(d-3)(2d+1)J^R_{42}(-\delta,0)\vecq{q}{2} +J^R_{42}(-\delta,\delta)\vecq{q}{2} 
\enh
-(d-3)(d-2)(d+1)J^R_{41}(-\delta,0) -(d+1)J^R_{41}(-\delta,\delta)
\Big]g^4 + 2\Big[ 
(d-3)q_0J^S_{11}(0)\vecq{q}{2} -q_0J^T_{11}(-\delta)\vecq{q}{2} +3(d-3)q_0J^S_{22}(0)\vecq{q}{2}
\enh
 -3q_0J^T_{22}(-\delta)\vecq{q}{2} +2(d-3)J^S_{24}(0)\vecq{q}{2} -2J^T_{24}(-\delta)\vecq{q}{2} +2(d-3)q_0J^S_{32}(0)\vecq{q}{2} -2q_0J^T_{32}(-\delta)\vecq{q}{2} 
 +2(d-3)J^S_{33}(0)\vecq{q}{2} 
  \enh
 -2J^T_{33}(-\delta)\vecq{q}{2}  -(d-3)(d-2)q_0J^S_{21}(0) -q_0J^S_{21}(\delta) + (d-1)q_0J^T_{21}(-\delta) -2(d-3)(d-2)q_0J^S_{31}(0) -2q_0J^S_{31}(\delta) 
 \enh
 +2(d-1)q_0J^T_{31}(-\delta) -2(d-3)(d-2)J^S_{34}(0) -2J^S_{34}(\delta) +2(d-1)J^T_{34}(-\delta)
\Big]g^2 
+ q_0^2J^F_0 +4q_0^2J^F_{11} +4q_0^2J^F_{21} +4J^F_{22}
\bigg\} g^{\mu\nu}
\enh
+ im_1 m_2 \frac{g^2}{2f^4} \bigg\{ \frac{-5}{2}\Big[ 
-(d-3)J^R_{22}(-\delta,0)\vecq{q}{2} + J^R_{22}(-\delta,\delta)\vecq{q}{2} -2(d-3)J^R_{32}(-\delta,0)\vecq{q}{2} +2J^R_{32}(-\delta,\delta)\vecq{q}{2} 
\enh
-(d-3)J^R_{43}(-\delta,0)\vecq{q}{2} +J^R_{43}(-\delta,\delta)\vecq{q}{2} +(d-3)J^R_{21}(-\delta,0) -J^R_{21}(-\delta,\delta) +(d^2-9)J^R_{31}(-\delta,0) 
-(d-3)J^R_{31}(-\delta,\delta) 
\enh
+(d^2-9)J^R_{42}(-\delta,0) -(d+3)J^R_{42}(-\delta,\delta)
\Big]g^2 
+(d-3)q_0J^S_{11}(0) -q_0J^S_{11}(\delta) +3(d-3)q_0J^S_{22}(0) -3q_0J^S_{22}(\delta) 
\enh
+2(d-3)J^S_{24}(0) -2J^S_{24}(\delta) +2(d-3)q_0J^S_{32}(0) -2q_0J^S_{32}(\delta) +2(d-3)J^S_{33}(0) -2J^S_{33}(\delta)
\bigg\}q^\mu q^\nu,
\\
\mathcal{K}^{I=1,\mu\nu}_{2\pi(2)}(p) =&im_1 m_2 \frac{-5(d-3)g^4}{4f^4}J^R_{21}(-\delta,-\delta)\Big[ \vecq{p}{2}g^{\mu\nu} +p^\mu p^\nu \Big].
	\end{align}
\end{widetext}

\bibliography{ref}

\end{document}